\documentclass{jfm_ed}

\usepackage{graphicx}
\usepackage{natbib}
\usepackage{cancel}
\usepackage[export]{adjustbox}

\ifCUPmtlplainloaded \else
  \checkfont{eurm10}
  \iffontfound
    \IfFileExists{upmath.sty}
      {\typeout{^^JFound AMS Euler Roman fonts on the system,
                   using the 'upmath' package.^^J}%
       \usepackage{upmath}}
      {\typeout{^^JFound AMS Euler Roman fonts on the system, but you
                   dont seem to have the}%
       \typeout{'upmath' package installed. JFM.cls can take advantage
                 of these fonts,^^Jif you use 'upmath' package.^^J}%
      }
  \else
  \fi
\fi

\ifCUPmtlplainloaded \else
  \checkfont{msam10}
  \iffontfound
    \IfFileExists{amssymb.sty}
      {\typeout{^^JFound AMS Symbol fonts on the system, using the
                'amssymb' package.^^J}%
       \usepackage{amssymb}%

      }{}
  \fi
\fi

\ifCUPmtlplainloaded \else
  \IfFileExists{amsbsy.sty}
    {\typeout{^^JFound the 'amsbsy' package on the system, using it.^^J}%
     \usepackage{amsbsy}}
    {}
\fi

\newsavebox{\astrutbox}
\sbox{\astrutbox}{\rule[-5pt]{0pt}{20pt}}

\usepackage{xcolor}

\newcommand{\be}{\begin{equation}}
\newcommand{\ee}{\end{equation}}
\newcommand{\bea}{\begin{eqnarray}}
\newcommand{\eea}{\end{eqnarray}}

\newcommand{\ba}{\begin{array}}
\newcommand{\ea}{\end{array}}

\newcommand{\xD}{{\bf D}}

\newcommand{\ez}{{\bf e}_z}

\newcommand{\xu}{{\bf u}}
\newcommand{\xq}{{\bf q}}

\newcommand{\Pe}{\mbox{Pe}}
\newcommand{\Ri}{\mbox{Ri}}

\renewcommand{\r}{s}

\def\Var{\mathop{\rm Var}}

\newcommand\underrel[2]{\mathrel{\mathop{#2}\limits_{#1}}}

\title[Testing gyrotactic swimmer dispersion in pipes]{Gyrotactic swimmer dispersion in pipe flow: testing the theory}

\author[O. A. Croze,
R. N. Bearon and M. A. Bees]
{Ottavio A. Croze$^1$%
  \thanks{Email address for correspondence: oac24@cam.ac.uk},\ns
Rachel N. Bearon$^2$
and Martin A. Bees$^3$}

\affiliation{$^1$Cavendish Laboratory, University of Cambridge\\ Cambridge, CB3 0HE, UK\\[\affilskip]
$^2$Department of Mathematical Sciences, University of Liverpool,\\ Liverpool L69 7ZL, UK\\[\affilskip]
$^3$Department of Mathematics, University of York, York YO10 5DD, UK}

\pubyear{2017}
\volume{816}
\pagerange{481--506}
\date{6 May 2016; revised 7 Feb 2017; accepted 9 Feb 2017}
\begin{document}

\maketitle

\begin{abstract}
Suspensions of microswimmers are a rich source of fascinating new fluid mechanics. Recently we predicted the active pipe flow dispersion of gyrotactic microalgae, whose orientation is biased by gravity and flow shear. Analytical theory predicts that these active swimmers disperse in a markedly distinct manner from passive tracers (Taylor dispersion). Dispersing swimmers display nonzero drift and effective diffusivity that is non-monotonic with  P{\'e}clet number. Such predictions agree with numerical simulations, but hitherto have not been tested experimentally. Here, to facilitate comparison, we obtain new solutions of the axial dispersion theory accounting both for swimmer negative buoyancy and a local nonlinear response of swimmers to shear, provided by two alternative microscopic stochastic descriptions. We obtain new predictions for suspensions of the model swimming alga {\it Dunaliella salina}, whose motility and buoyant mass we parametrise using tracking video microscopy. We then present a new experimental method to measure gyrotactic dispersion using fluorescently stained {\it D. salina} and provide a preliminary comparison with predictions of a nonzero drift above the mean flow for each microscopic stochastic description. Finally, we propose further experiments for a full experimental characterisation of gyrotactic dispersion measures and discuss implications of our results for algal dispersion in industrial photobioreactors. 
\end{abstract}

\begin{keywords}
Authors should not enter keywords on the manuscript, as these must be chosen by the author during the online submission process and will then be added during the typesetting process (see http://journals.cambridge.org/data/\linebreak[3]relatedlink/jfm-\linebreak[3]keywords.pdf for the full list)
\end{keywords}

\section{Introduction}\label{sec:intro}

The behaviour of microbial biofluids is important to a broad range of diverse fields from medicine and biotechnology to biogeochemistry, aquatic ecology and climate science. When suspended cells self-propel the contrast with passive particle hydrodynamics is particularly dramatic. For example, dilute suspensions of swimming bacteria, algae and ciliates display self-concentration and hydrodynamic instabilities \citep{PedleyKessler92}. The dilute and concentrated suspension rheology of swimmers is also of much interest: for example, swimming algae and bacteria can enhance or reduce effective suspension viscosity, respectively \citep{Marchettietalreview13}. Many motile microorganisms respond to their environment and migrate towards nutrients (chemotaxis) or light (phototaxis). Intriguingly they also respond to flow gradients. In this work we focus on gyrotaxis, the passive orientational bias on swimming microalgae due to a combination of viscous and gravitational torques, but the results may be generalized to other taxes or combinations thereof. In downwelling pipe flow, gyrotactic swimmers self-focus in structures known as plumes \citep{Kessler85}. Though the discovery of gyrotaxis in algae is now over 30 years old, many open questions remain. Here, we shall address the dispersion of gyrotactic swimmers, and how well experiments can be described by existing theoretical models of gyrotaxis. 

The classic papers by G. I. Taylor and R. Aris considered the dispersion of passive solutes in laminar pipe flow, and showed that the cross-sectional averaged axial solute dispersion can be mapped to an effective diffusion, with diffusivity $D_e=D_m(1+\rm{Pe}^2/48)$ \citep{Taylor53,Aris56}. Taylor tested this prediction experimentally himself \citep{Taylor53}, as well as its extension to turbulent pipe flow \citep{Taylor54a}. Recently, we revisited the Taylor-Aris analytical theory for laminar flow to describe the dispersion of swimmers in confined flow \citep{BeesCroze10} and used it to predict the dispersion of gyrotactic swimmers in laminar and turbulent flows \citep{BearonBeesCroze12, Crozeetal13}. The  \cite{BeesCroze10} theory is general and applicable to any microswimmer. It requires as input the mean response to flow of swimmers from stochastic microscopic models. We have applied it to gyrotactic microalgae, where stochastic models of self-propelled spheroids in flows have been developed \citep{PedleyKessler90, PedleyKessler92, HillBees02, ManelaFrankel03, BearonBeesCroze12}, comparing predictions from analytical theory with individual based numerical simulations \citep{Crozeetal13} and numerical solutions of the model advection-diffusion equation \citep{BearonBeesCroze12}. 

The present work on dispersion was inspired by the notion that biotechnologically useful microalgal species, such as the $\beta$-carotene producing biflagellated alga {\it Dunaliella salina} are cultured in pipe flow within industrial tubular photobioreactors \citep{Crozeetal13, BeesCroze14}. As we shall discuss below, gyrotaxis at the individual cell level dramatically modifies the population-scale behaviour of suspensions of biflagellates such as {\it D. salina}, and this has important consequences for the engineering design of photobioreactors, for both laminar and turbulent flow regimes \citep{Crozeetal13}. Evidence is also mounting that gyrotactic behaviour is relevant to environmental flows. \cite{Durhametal09} proposed gyrotactic trapping to be one of the mechanisms leading to the formation of oceanic thin layers \citep{DurhamStocker12}. Theory and simulation also predict similar effects in turbulent flows, where gyrotactic swimmers `unmix' in downwelling regions of turbulent flows \citep{Crozeetal13,Durhametal13}, or even due to acceleration in strong turbulence \citep{DeLilloetal14}. Furthermore, the study of bioconvection \citep{PedleyKessler92, HillPedley05}, an instability driven by upswimming and gyrotaxis, has recently seen some interesting developments. Weak shear flow has been shown to distort but not destroy bioconvection patterns \citep{CrozeAshrafBees10, HwangPedley14}. Light from below can frustrate the patterns \citep{WilliamsBees11, WilliamsBees11b}, while in microchannels, light from the side gives rise to horizontal accumulation \citep{Garciaetal13}. The cell excess density (the difference between cell and surrounding medium density) can give rise to instabilities in gyrotactic plumes known as blips \citep{Kessler86,PedleyKessler92,DennisenkoLukashuck07}. The existence of blips as a function of imposed flow and swimmer concentration can be predicted theoretically using linear stability analysis \citep{HwangPedleyBlips14}. Even in the absence of a gravitational torque, algae can display peculiar behaviour, such as limit cycle oscillations in pipe flow \citep{ZoettlStark12}. Finally, while it is well known that algae swim with a helical trajectory, only recently have the mechanics been addressed \citep{Bearon13}, qualitatively explaining resonant alignment in oscillatory shear flow \citep{Hopeetal16}.

Quantitative comparison between mathematical theory and experiment is essential in fluid mechanics. G.I.~Taylor excelled and delighted in this comparison \citep{CrozePeaudercef16}. Without it theory has no constraint and experiments no mechanism. Furthermore, a quantitatively tested theory is of greater engineering use. In this paper, we set out to test the theory of gyrotactic swimmer dispersion experimentally for vertical pipe flow. In section \ref{sec:theory} we summarise the theory of axial cell dispersion, which links the statistical individual-level response to flow, predicted from two distinct microscopic stochastic models, with the population-level transport measures: the effective axial drift and diffusivity. In the process of experiment-theory iteration, we identified that cell concentration affected dispersion. Therefore we explore how algal cell negative buoyancy perturbs the flow field from simple Poiseuille flow, which further modifies the radial distribution of cells, altering axial dispersion. Dispersion with negative buoyancy had previously only been considered for a linearised response to flow \citep{BeesCroze10}. Here, we employ expressions for the nonlinear response to flow predicted by microscopic stochastic descriptions. In section \ref{sec:experiments} we describe experimental methods for cell culture, fluorescent staining and imaging, as well as data analysis and comparison with theoretical predictions. We also explain the challenges due to having to maximise fluorescent image contrast while avoiding blip instabilities. In spite of these issues, we establish a robust experimental protocol to study gyrotactic dispersion and obtain preliminary results allowing one to compare predicted and measured axial drift for various average cell concentrations. Finally, in section \ref{sec:discussion} we propose further experiments to quantify dispersion and discuss the broader relevance of our results for growing algae in photobioreactors.

\section{Theory}\label{sec:theory}

In this paper, we test theoretical predictions of biased microswimmer dispersion with experimental data.  To enable a direct comparison we derive new results from the general axial dispersion framework \citep{BeesCroze10, BearonBeesCroze12, Crozeetal13} to include both the effect of swimmer negative buoyancy and the complete nonlinear response of swimmers to shear. We summarise the theory below, referring the reader to the above papers for further mathematical details. 

Consider a suspension of gyrotactic swimming algae (such as {\it D. salina}) subject to imposed flow in a pipe. Gyrotactic swimmers actively cross streamlines and focus in downwelling flows, which is at the root of their peculiar {\it non-Taylor} dispersion. Following \citet{PedleyKessler92} and the assumptions discussed therein, an incompressible, dilute suspension of gyrotactic swimming cells in an axial flow in a tube can be described by the equations
\bea
&& \frac{\partial{\bf u}}{\partial t}=\frac{1}{\rho}\left[-\nabla p+ \mu \nabla^2 {\bf u}  + n v_c \Delta \rho {\bf g}\right]\label{eq:NS};\\
&& \frac{\partial n}{\partial t}=-\nabla \cdot \left[ n \left ( {\bf u } + {\bf q}V_s\right) - \frac{V_s^2}{d_r}{\bf  D} \cdot \nabla  n \right],\label{eq:cells}
\eea
where (\ref{eq:NS}) is the Navier-Stokes equation for the pipe flow {\bf u}, incorporating a driving pressure gradient term (with pressure, $p$), a Newtonian viscous stress (medium viscosity, $\mu$, approximately equal to that of water) and a negative buoyancy term (cell volume, $v_c$; density excess, $\Delta \rho$; gravitational acceleration, ${\bf g}$; cell concentration, $n$). Negative buoyancy causes cell accumulations to sink, which can result in instabilities such as blips \citep[][]{HwangPedleyBlips14}; see sections \ref{sec:experiments} and \ref{sec:blips} below.   Equation (\ref{eq:cells}) expresses conservation of cells, with advection contributions from flow and swimming (velocity, ${\bf q}V_s$; mean orientation, ${\bf q}$; and speed,  $V_s$) and swimming diffusion (with anisotropic diffusivity tensor, {\bf  D}, and magnitude $V_s^2/d_r$, where $d_r$ is the rotational diffusivity). No-slip and no-flux boundary conditions are applied at the tube wall.

Equations (\ref{eq:NS}) and (\ref{eq:cells}) describe the dynamics of a suspension of gyrotactic swimming algae. As mentioned in Section~\ref{sec:intro}, such suspensions self-concentrate into plumes; cells focus in a plume at the center of a vertically aligned tube and drive axial flow. However, we are free to label (or dye) locally some of these cells (which we call a `slug') without affecting the cell concentration profile or the flow.  
The theory presented here describes the dispersion of a labelled slug of algae in the flow associated with an existing plume. The dynamics of the moments of a distribution of labelled cells are evaluated, providing information on axial drift and diffusion.
We summarise the derivation of the main results as follows. Steady axisymmetric solutions will be obtained for  (\ref{eq:NS}) and (\ref{eq:cells}) (steady plume solutions). Then, moment equations from (\ref{eq:cells}) will be used to evaluate dispersion measures, using the mean cell response to flow from two alternative stochastic microscopic models. Finally, the predictions will be analysed for direct comparison with experiments.

\subsection{The steady plume background}

First, consider the background flow within which the slug of cells will disperse, provided by the steady solutions to (\ref{eq:NS}) and (\ref{eq:cells}) (a steady plume). As in \citet{BeesCroze10}, we assume a plume with no blips or varicose instabilities has formed in a long pipe, so that the dynamics are translationally invariant along the axial direction $z$. We further assume axisymmetric solutions for the cell concentration $n(r)$ and flow deviation about the mean $\chi(r)$. In cylindrical polars, the flow field is $\xu(r) = u(r) \ez = U[1+\chi(r)] \ez$, with $U$ the mean flow. We nondimensionalize lengths by the pipe radius,  $\hat{x}=x/a$, times by $\tau_d=a^2 d_r/V_s^2$, the time to diffuse across it, such that $\hat{t}=t/\tau_d$, and concentrations by the background mean cell concentration $n_{av}$, such that $R_0^0=n/n_{av}$ (this notation is chosen for consistency with \citet{BeesCroze10}; see next section). Equations (\ref{eq:NS}) and (\ref{eq:cells}) then become (omitting hats for clarity)
\bea
&&\frac{1}{r} \frac{d}{dr} \left( r \frac{d \chi}{dr}  \right) =  P_{z} - \Ri\,R_0^0;\label{eq:NSd}\\
&& \frac{d R_0^0}{dr} = \beta \frac{q^r}{D^{rr}} R_0^0,\label{eq:cellsd}\label{eq:R00}
\eea
subject to no-slip and no-flux boundary conditions at the tube surface, and the integral constraints 
\be\label{eq:chiR00norm}
\overline{\chi}=2\int_0^1 r \chi(r) dr=0;\mbox{~~}\overline{R_0^0}=2\int_0^1 r R_0^0(r) dr=1,
\ee
where we denote cross-sectional averages by overbars: $\overline{f}(z,t)=\frac{1}{\pi}\int_0^{2\pi}\int_0^1f(r,\theta,z,t) \: r\, dr\, d\theta$.  If $f$ is independent of $\theta$ then $\overline{f}(z,t) = 2\int_0^1f(r,z,t) \, dr$. Equations (\ref{eq:chiR00norm}) follow from the definition of $\chi$ and normalisation of the cross-sectional average concentration. In (\ref{eq:NSd}) and (\ref{eq:cellsd}), we have defined the dimensionless parameters
\be
P_z = \frac{1}{\mu}\frac{d p}{dz}\frac{a}{U};\mbox{~~}\Ri = \frac{v_c \,\Delta \rho\,g}{\mu}\frac{a^2}{U}n_{av};\mbox{~~}\beta = \frac{a d_r}{V_s}.
\ee
where $P_z$ represents the dimensionless pressure gradient in the z-direction; $\Ri$ is a Richardson number quantifying the relative importance of buoyancy and viscous flow ($\mu$ is the dynamic viscosity); $\beta$ is the swimming P{\'e}clet number gauging the relative importance of advection by swimming to swimming diffusion (see also equation (\ref{CC1})). Alternatively, $\beta=1/{\rm Kn}$, where ${\rm Kn}=(V_s d_r^{-1})/a$ is a swimming Knudsen number: the ratio of the cell mean free path to the tube radius. 
To close equations (\ref{eq:NSd}) and (\ref{eq:cellsd}), we require the radial components of the swimming direction, $q^r=q^r(\chi^\prime)$, and diffusivity tensor, $D^{rr}(\chi^\prime)$, where $\chi^\prime\equiv d \chi/d r$. These are known functions of shear obtained from stochastic models of gyrotactic response to flow and are discussed in section \ref{sheartrans}.

\subsection{Slug dispersion}

Next, we consider the dynamics of a slug of dyed cells dispersing within the plume flow background provided by the solutions of equations (\ref{eq:NSd}) and (\ref{eq:cellsd}). A suspension of dyed swimmers replace cells in the pipe within an existing plume, forming a dyed slug. The dyed swimmers are identical to the background cells in the plume and the slug dynamics are also governed by (\ref{eq:cells}). However, since the cells are initially localised and subsequently disperse the solutions are no longer independent of the axial coordinate or time. The slug concentration, denoted by $n_s$ (nondimensionalized by $n_{av}$ as above, but not yet cross-sectionally averaged) is governed by
\be\label{eq:cellsd-slug}
\frac{\partial n_s}{\partial t} = \nabla \cdot \left( \xD \cdot \nabla n_s \right) - \Pe (1+\chi) \frac{\partial n_s}{\partial z}
- \beta \nabla \cdot \left( n_s \xq \right), \label{CC1}
\ee
where 
\be
\Pe = \frac{U a d_r}{V_s^2}
\label{eq:Pe}
\ee 
is the flow P{\'e}clet number quantifying the relative importance of advection by the flow to swimming diffusion.

Except for a few special cases, equation (\ref{CC1}) is not amenable to analytical solution. It can be solved numerically using a spatially adaptive finite element method, as described in \cite{BearonBeesCroze12}. This method allows to probe the transient dispersion of a slug and the approach to steady state. To characterise dispersion it is convenient to transform to a frame of reference moving with the mean flow: $z \to z-\Pe\,t$ and consider axial moments of concentration. The $p$-th radially varying moment is defined as $c_p(r,t) = \int^{+\infty}_{-\infty} z^p n_s(z,r,t) dz$, and its cross-sectional average as $m_p(t) = \overline{c_p}=2\int^1_0 r c_p(r,t) dr$. We define the transient drift and diffusivity of the dispersing slug as:
\be\label{driftDetrans}
\Lambda_0(t)\equiv\frac{d}{d t}m_1(t); \mbox{~~~~~~~~~} D_e(t)\equiv\frac{1}{2}\frac{d}{d t}\left[m_2(t)-m_1^2(t)\right].
\ee
An alternative to solving equation (\ref{CC1}) is to solve for the axial moments $c_p(r,t)$ and cross-sectionally averaged moments $m_p(t)$ \citep{BeesCroze10}. The theory then allows one to predict that at long times the drift, $\Lambda_0$, and effective diffusivity, $D_e$, of a dyed slug of swimmers are given by (with the notation $f^\prime=df/dr$):
\bea
\Lambda_0 &\equiv&\lim_{t\to\infty}\Lambda_0(t)=  -\overline{D^{rz}R_0^{0\prime}}
+ \overline{\left( \Pe \chi + \beta q^{z} \right) R_0^0}, \label{drifteff} \\
D_{e}&\equiv&\lim_{t\to\infty}D_e(t)= - \overline{D^{rz} g^{\prime}} 
+ \overline{ \left( \Pe \chi +\beta q^{z} -\Lambda_0\right) g}+ \overline{D^{zz} R_0^0}, \label{Deff}
\eea
where $R_0^0(r)$ is the background normalised concentration of cells from equations (\ref{eq:NSd}) and (\ref{eq:cellsd}), and
\be \label{eq:g}
g(r)=R_0^0 \int^r_0  \left( \frac{D^{rz}(\r)}{D^{rr}(\r)}- \frac12 \frac{\Lambda_0^*(\r), 
- \Lambda_0 m_0^*(\r)}{\r D^{rr}(\r)R_0^0(\r)}  \right) d\r\label{eq:fr}
\ee
with
\be \label{eq:Lam0starm0star}
\Lambda_0^*(r) = 2 \int^r_0  \r\left( -D^{rz} {R_0^0}^{\prime} + \left( \Pe \chi + \beta q^x \right) R_0^0  \right)d\r;  \mbox{~~~} 
m_0^*(r) = 2 \int^r_0 \r R_0^0(\r) d\r.
\ee
%

The function $g(r)$ (first axial moment) gauges how the center-of-mass (mean axial position) of a dyed slug varies radially. This controls the effective diffusivity, $D_e$, of the slug together with the cell concentration distribution $R_0^0(r)$ (zeroth moment), which prescribes the slug drift, $\Lambda_0$. The function $\Lambda_0^*(r)$ in equation (\ref{eq:Lam0starm0star}) is the radially averaged slug cell flux (due to anisotropic diffusion, flow and swimming) up to radius $r$. Evaluated at the wall ($r=1$) this provides the full non-dimensional drift: $\Lambda_0^*(1)=\Lambda_0$. Similarly, $m_0^*(r)$ represents the radially averaged slug cell concentration up to $r$. At the wall this gives $m_0^*(1)=1$ (equivalent to equation  (\ref{eq:chiR00norm})), which is the total nondimensional number of cells reflecting cell conservation.

To obtain experimental predictions from (\ref{drifteff}) and (\ref{Deff}) we require functional forms for the flow about the mean, $\chi$, the average swimming direction, ${\bf q}$, and diffusivity tensor, ${\bf D}$. Expressions for ${\bf q}$ and ${\bf D}$ for gyrotactic algae can be obtained from microscopic stochastic models, as detailed in the next section. If we neglect reciprocal coupling of the flow to the cells via cell negative buoyancy,
we have Poiseuille flow, and expressions for ${\bf q}$ and ${\bf D}$ are given in \citep{BearonBeesCroze12}. Accounting for reciprocal coupling requires solving equations (\ref{eq:NSd}) and (\ref{eq:cellsd}) for $\chi$ and $R_0^0$, subject to the constraints (\ref{eq:chiR00norm}). From this we can make predictions for dispersion using the integrals (\ref{drifteff}) and (\ref{Deff}). Numerically it is easier to solve the equivalent ODE system (see next section and Appendix \ref{sec:appEblips} for details).

\subsection{Shear dependence of transport, model parameters and solution \label{sheartrans}}

The components of the mean cell orientation, ${\bf q}$, and diffusivity, ${\bf D}$, in equations (\ref{CC1}) and (\ref{drifteff}-\ref{eq:fr}) gauge average gyrotactic cell response to shear flow. They can be obtained from the orientational moments of the probability density function (PDF) for cell orientation and possibly position in a flow from stochastic continuum descriptions. As in previous work, we consider two such models, denoted F and G (described as Fokker-Planck and Generalised Taylor Dispersion models in the literature). In model F \citep{PedleyKessler90} the cell PDF depends only on orientation, while in G \citep{HillBees02, ManelaFrankel03, BearonBeesCroze12} it depends on position and orientation. Models F and G were originally derived for unbounded linear shear flows \citep{HillBees02, Beesetal98} (see appendix for (\ref{sec:appCmicro}) a model equation summary. However, they also provide a good approximation for the swimmers pipe (radius $\sim a$) flow, if the flow shear is linear on the scale of a cell (radius $r_c$): $r_c\ll a$ \citep{BearonBeesCroze12}. A separate requirement is that the cell diffusion is not in the Knudsen regime, $\rm{Kn}(=\beta^{-1})\sim1$, in which the cell random walk strongly affected by walls. Using parameters from Table \ref{ModPar} and $r_c\sim R$, where $R\sim10^{-4}$ cm is the cell hydrodynamic radius (see appendix (\ref{sec:appCmicro}), we see that  $r_c/a\sim 10^{-4}$ and $\rm{Kn}\sim 0.1$, so that both conditions are well-satisfied in our experiments. In both models F and G, a key dimensionless parameter is the stochasticity parameter $\lambda=1/(2 d_r B)$, expressing the relative importance of random reorientation from rotational diffusion $d_r$ to reorientation by gravity at a rate $1/(2B)$. The PDF can be solved using a Galerkin method for a given value of $\lambda$. This has been achieved previously for ${\it C. augustae}$ \citep{Beesetal98, BearonBeesCroze12}. Here, we have obtained solutions for {\it D. salina} (see below and Appendix \ref{sec:appAqrDrr}), allowing the evaluation of the functions $q^i(\sigma)$ and $D_m^{ij}(\sigma)$, where $m=$ F or G and $\sigma=-\chi^\prime\Pe/(2 \beta^2)$, the dimensionless shear rate. The model subscript in the average orientation components has been omitted intentionally: the predictions are the same for both models. The difference is in the diffusivity components, reflecting the different physics in the models.  In F the cell orientation distribution alone determines spatial dispersion, but in G both orientation and position (leading to differential advection by swimming and flow) contribute to dispersion via spatial moments. 
Indeed, predictions for diffusion from the F model are strictly valid at low shear rates, and break down for large shear rates (see \citet{Crozeetal13}). The G model is valid for all shear rates, provided the shear flow can be locally approximated as linear. The description is appropriate for the Poiseuille flows considered here, but not straining flows \citep{BearonHazelThorn11}. Both models provide approximations to the full Schmoluchowski advection-diffusion equation for the cell PDF \citep{DoiEdwards}, which could  be solved numerically directly for the cell PDF\citep{SaintillanShelley08}. However, this route does not allow one to take advantage of the simplification and analytical results that have been obtained for the F and G models \citep{PedleyKessler92, BearonBeesCroze12}. 

The dispersion measurements described in Section \ref{sec:exp} provide the first experimental test of the validity of the microscopic stochastic models described above. The dispersion model was parametrised from independent tracking video microscopy measurements, carried out on suspensions of {\it D. salina} in vertically oriented capillaries. The model parameters thus obtained and used in predictions are summarised in Table \ref{ModPar}. The stochasticity parameter $\lambda$ from tracking was used to determine the specific forms of $q^i(\sigma)$ and $D_m^{ij}(\sigma)$ for {\it D. salina} (Appendix \ref{sec:appAqrDrr}). Tracking microscopy was also used to quantify the settling of heat-immobilised cells, which allows to infer the buoyant mass $\Delta \rho\,v_c$ of {\it D. salina} cells. More details of the tracking measurements and error estimation can be found Appendix \ref{sec:appCmicro}. The model was solved numerically using MATLAB (Mathworks, Natick, MA, USA). As detailed in Appendix \ref{sec:appDnumsol}, it is easier to recast the dispersion prediction (integral equations (\ref{drifteff}-\ref{Deff}) and dependent equations) as a boundary value ODE problem in which the coupled flow, cell concentration and dispersion are evaluated together. It should be noted that the dimensionless parameters of the dispersion and microscopic models are not all independent. In particular, the Richardson number depends on Pe and $\beta$: $\Ri=\gamma \frac{\beta^3}{\rm{Pe}}$. The coupling of cells and flow thus depends on imposed flow and the coupling strength $\gamma\equiv \frac{v_c\,\Delta \rho g}{\mu}\frac{V_s}{d_r^2}n_{av}$, which we can control by changing the background mean cell concentration $n_{av}$.




%
\begin{table}
\begin{center}
\begin{tabular}{l l l l}
Mean swimming speed & $V_s$	 & $(6.27 \pm 0.04){\times} 10^{-3}$ cm s$^{-1}$ & this work\\
Bias parameter & $\lambda$	 & $0.21\pm0.05$ & this work\\
Gravitactic reorientation time & $B$	 & $10.5\pm1.3$ s & this work\\
Rotational diffusivity & $d_r=(2 B \lambda)^{-1}$	 & $0.23 \pm 0.06$ s$^{-1}$ & this work\\
Excess buoyant mass & $\Delta\rho\,v_c$ & $(0.92\pm0.36){\times} 10^{-11}$ g & this work\\
Tube radius & $a$ & $0.35\pm0.02$ cm & this work \\
Gravitational acceleration & $g$ & $9.8{\times}10^2$ cm s$^{-2}$ & \citep{HandChemPhys03}\\
Dynamic viscosity ($25^\circ$ C) & $\mu$ & $9{\times}10^{-3}$g cm$^{-1}$ s$^{-1}$ & \citep{Phillipsetal80}\\
Mean cell concentration & $n_{av}$ & $0.70$-$3.65{\times} 10^6$ cells cm$^{-3}$ & this work\\
Imposed flow rate & $\Phi$ & $100$-$400$ cm$^3$ h$^{-1}$ & this work\\
Mean flow speed & $U=\frac{\Phi}{\pi a^2}$ & $0.07$-$0.29$ cm s$^{-1}$ & this work\\\\ 
Swimming P{\'e}clet number & $\beta=\frac{a d_r}{V_s}$	 & 12.2  \\
P{\'e}clet number & $\Pe=\frac{U a d_r}{V_s^2}$	 & $280$-$560$\\
Cell-flow coupling parameter & $\gamma\equiv \frac{v_c\,\Delta \rho g}{\mu}\frac{V_s}{d_r^2}n_{av}$ 	 & $0.08$-$0.44$ \\
Richardson number & $\Ri=\gamma \frac{\beta^3}{\rm{Pe}}$ 	 & $0.26$-$2.85$ 
\end{tabular}
\caption{Dimensional and nondimensional parameter values and ranges used in the dispersion model predictions. The {\it D. salina} gyrotactic motility and buoyancy parameters were estimated from tracking videomicroscopy measurements (see text and Appendix \ref{sec:appAqrDrr}).} \label{ModPar} 
\end{center}
\end{table}
%


\subsection{Dispersion predictions}

Before turning to a comparison between dispersion theory and experiment, it is useful to consider predictions for the distribution of gyrotactic algae in pipe flow and the ensuing dispersion. Similar predictions have been obtained using the swimming parameters of {\it C. augustae} \citep{BearonBeesCroze12}, and also for channel geometries \citep{Crozeetal13}, comparing predictions for the F and G microscopic models. These studies assumed the coupling via buoyancy between cells and flow dynamics is negligible. Reciprocal coupling was considered by \cite{BeesCroze10}, but only with asymptotic results from the F model. Here, we present the first solutions of the dispersion equations with reciprocal coupling by buoyancy and nonlinear expressions from the stochastic microscopic descriptions F and G, and use them to investigate the dispersion of {\it D. salina}, the fully parametrised candidate species for experimental tests of the theory. We consider long-time dispersion, i.e. when axial moments are stationary in time. This is true when $t\gg 1/\zeta_1^2$, where $\zeta_1$ is the smallest positive eigenvalue arising in the moment equations for cell conservation; $\zeta_1({\rm Pe})$ is a monotonically increasing function of Pe, which is laborious to evaluate (see section 4 of \cite{BeesCroze10} for details). Simulations of gyrotactic dispersion in channels confirm that steady dispersion is reached faster for larger Pe and, further, that the drift (measured in this work) becomes steady faster than the diffusivity \citep{Crozeetal13}. For ${\rm Pe}=670$, close to the largest ${\rm Pe}=560$ value studied here, these simulations show that the drift is steady at $t\approx 0.5$. That is to say, restoring dimensions, at a time $t_c\approx\tau_d/2$, where recall $\tau_d=a^2 d_r/V_s^2$ is the swimming diffusivity timescale. Multiplying by the mean flow speed $U$, and recalling the definition of Pe (equation \ref{eq:Pe}), this implies dispersion will be steady beyond a distance $Z_c\approx {\rm Pe}\,a/2$ from the inoculation point (for general Pe, $Z_c\approx {\rm Pe}\,a/\zeta_1^2$). Thus, with $a=0.35$ cm (see Table \ref{ModPar}) we estimate that for the largest ${\rm Pe}=560$ used in our experiments, $Z_c\approx 100$ cm, the distance  downstream from dyed cell injection at which our imaging was carried out (see section \ref{sec:methods}). We can thus assume our experimental drift data is steady and it is reasonable to compare it with long-time dispersion predictions (see Figure \ref{fig:drift_thvsexp}).
\begin{figure}
\center{
\includegraphics[width=0.5\linewidth]{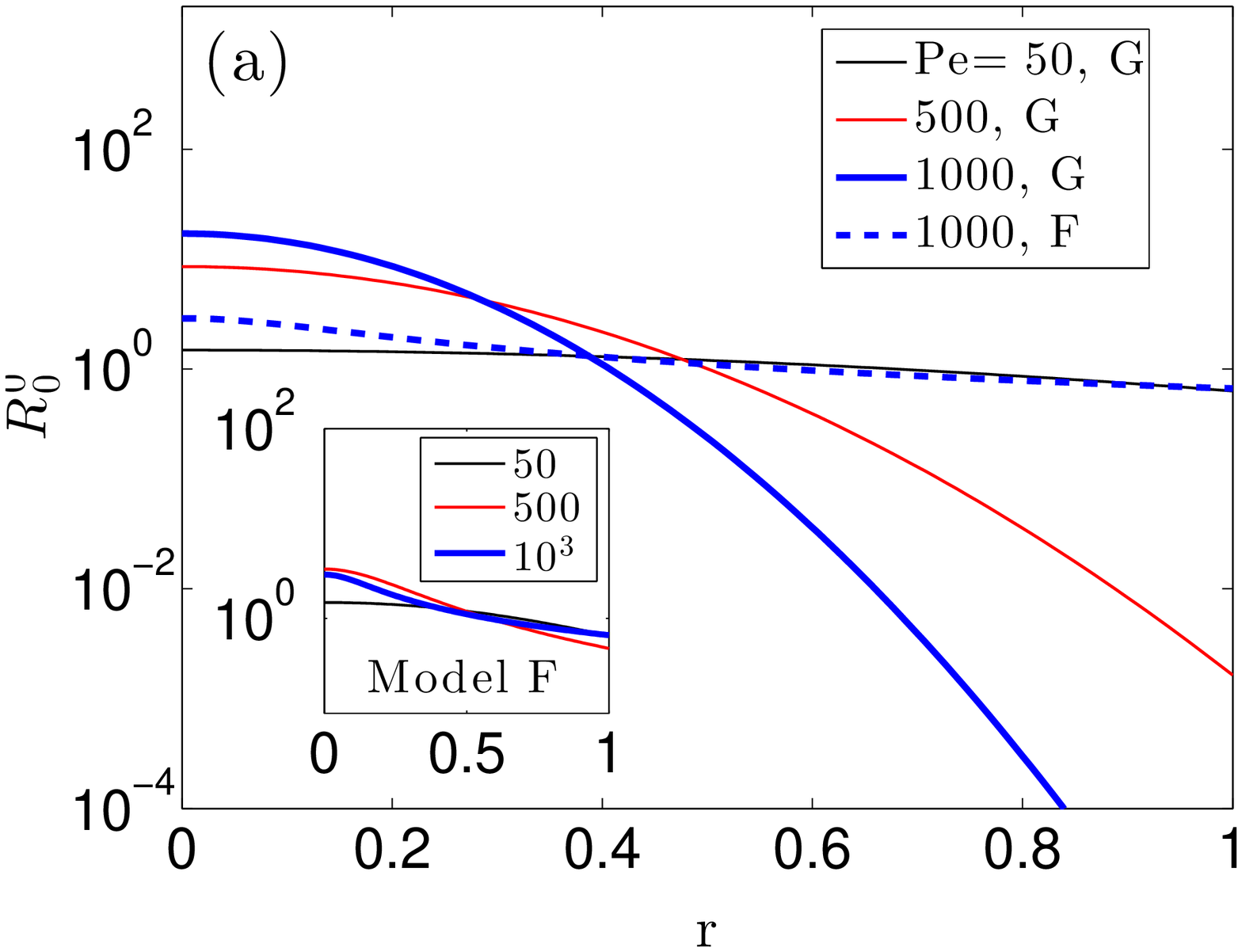}\\
\includegraphics[width=0.5\linewidth]{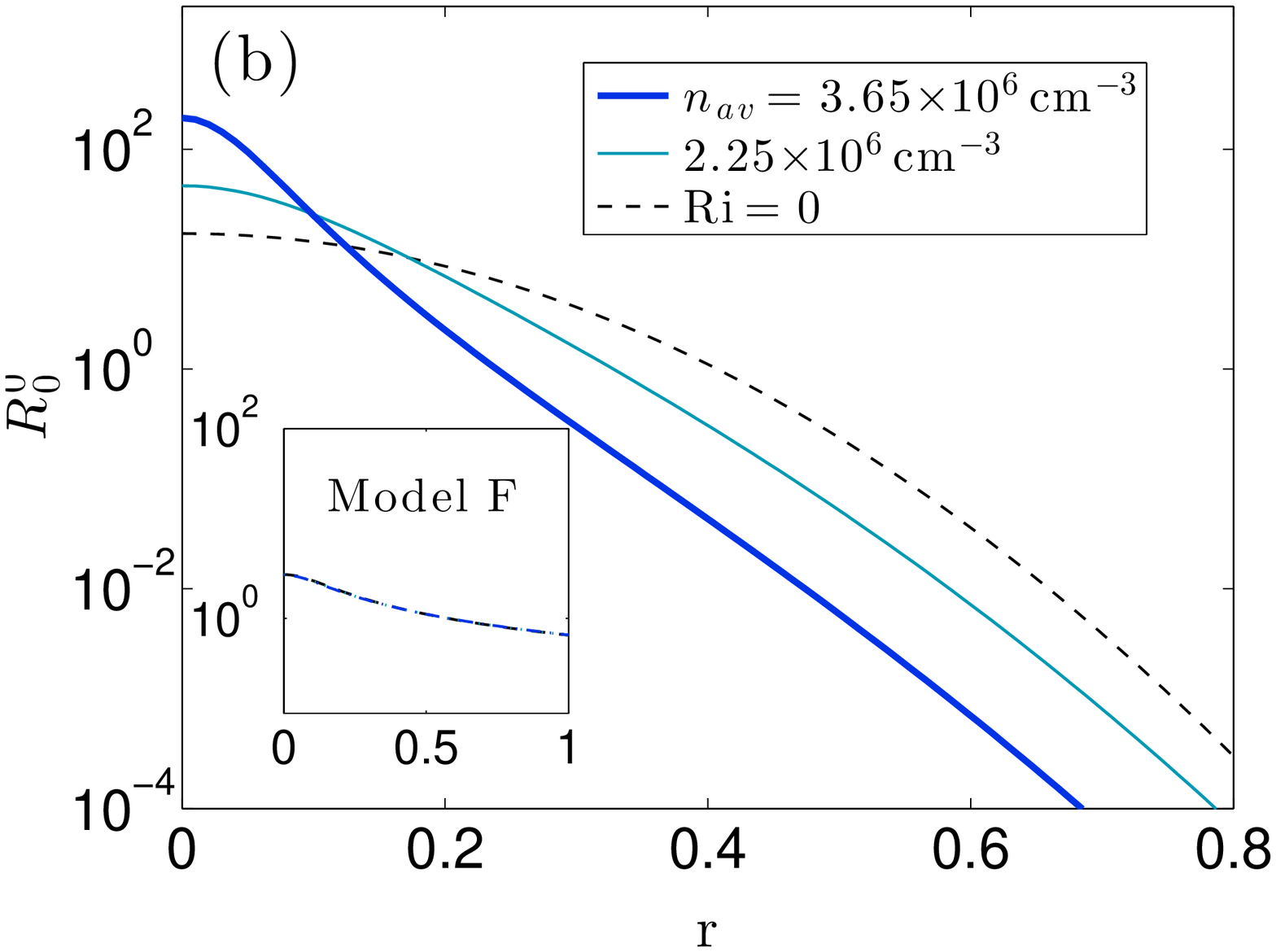}\\
\includegraphics[width=0.5\linewidth]{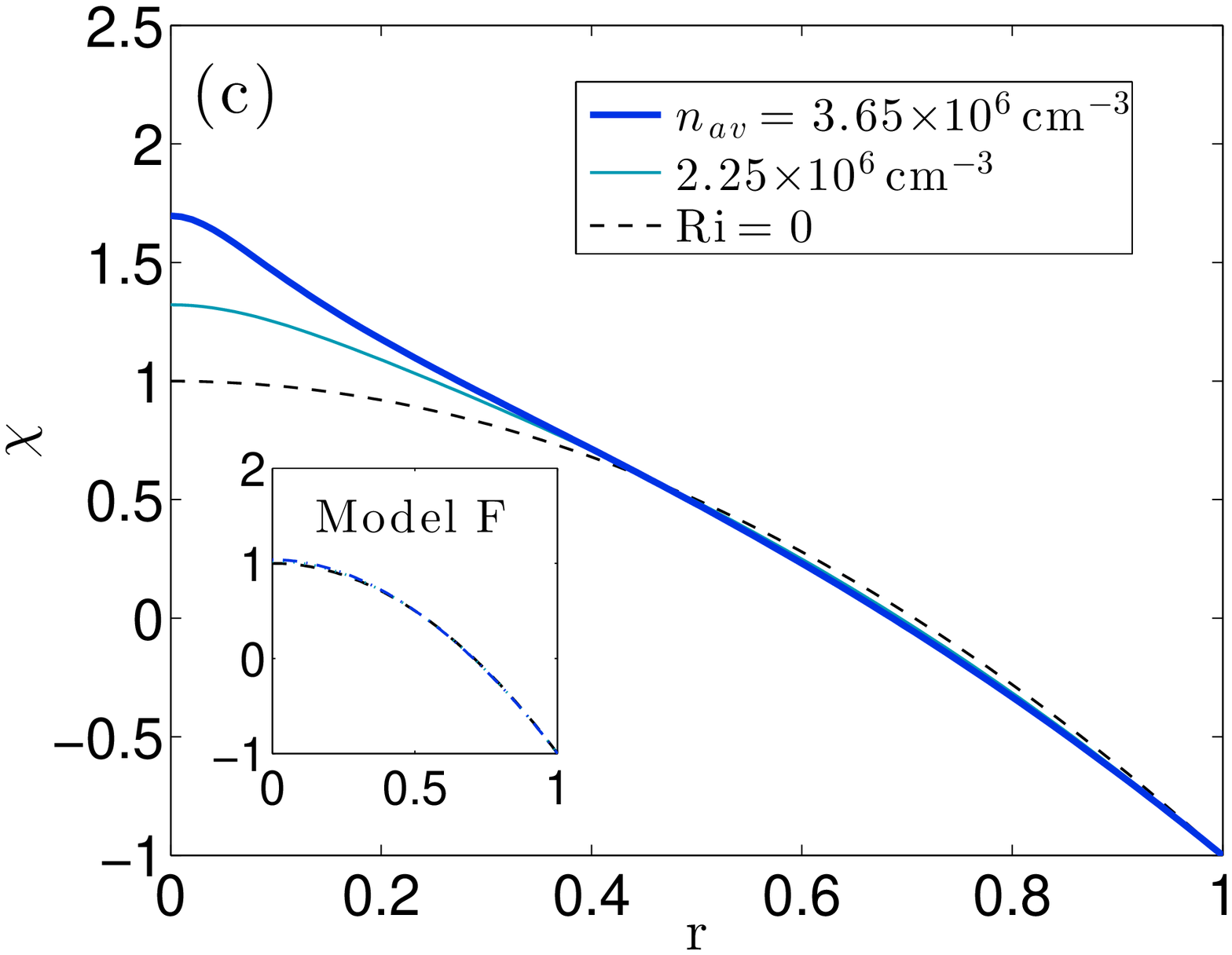}
}
\caption{(a) Dimensionless normalised cell concentration profiles for different values of P{\'e}clet number, Pe (from Equation \ref{eq:R00}) using the G model for neutrally buoyant cells (Ri$=0$, no cell-flow coupling). The distribution is increasingly focused with increasing Pe, unlike the F model predictions (inset). (b, c) Reciprocally coupled cell concentration and flow as a function of mean background cell concentration $n_{av}$ for the G model and Pe$=1000$. The coupling generates sharper focusing and distorts the flow from the parabolic Poiseuille shape. Insets: for the F model the coupling hardly affects the cell and flow distributions. Note that, since $V_s$ and $d_r$ are fixed, changing \Pe\, corresponds to changing the mean speed $U$ (or flow rate), as in our dispersion experiments.}\label{fig:cellconc}
\end{figure}

In figure \ref{fig:cellconc} we consider predicted cell distributions. For the case with no reciprocal coupling (neutral buoyancy; Richardson number, $\rm{Ri}=0$), shown in Figure \ref{fig:cellconc}a, both the F and G models predict gyrotactic focusing. However, the distributions are different: for the G model, as the P{\'e}clet number Pe is increased, the distribution goes from almost uniform to sharply focused (the width diminishes monotonically). On the other hand, as Pe increases the F model (Figure \ref{fig:cellconc}a, inset) displays a broader distribution (with a non-monotonic width progression). The differences between the two models, which are noticeable for Pe$=1000$, arise from the different average microscopic transport these models predict \citep{BearonBeesCroze12}. In model G, local cell dispersion depends on reorientation by the flow and differential advection due to the spatial PDF; in the F model the local spatial distribution is neglected (a good approximation only for weak flows). As a result, the models predict the same mean cell orientation $q^i$, but different diffusivities $D^{ij}$ as a function of dimensionless shear $\sigma$ \citep{HillBees02, ManelaFrankel03, BearonBeesCroze12}. The cell distribution is determined by the ratio of the radial components of these transport parameters, $q^r/D^{rr}$, as is clear from equation (\ref{eq:R00}). In model F, because $D^{rr}$ asymptotes to a constant value at large $\sigma$ \citep{BearonBeesCroze12}, gyrotactic advection in high shear (near the pipe walls) is weak compared to diffusion, so the cell distribution has significant tails at high Pe (see \ref{fig:cellconc}a, inset). In model G, on the other hand,  $D^{rr}\to 0$ at large $\sigma$ \citep{BearonBeesCroze12}: gyrotactic advection is strong compared to diffusion near the walls, and the distribution is sharp, with low cell concentration close to the walls. 

Next, we consider the distributions of cells and flow with the additional coupling via the cells' negative buoyancy (Figure \ref{fig:cellconc}b).  As discussed above, the strength of coupling $\gamma$ for a given Pe depends on the cell concentration $n_{av}$. In model G, for a given Pe increasing $n_{av}$  results in a sharper distribution near the center of the tube with respect to the uncoupled case. Concomitantly, cells that have accumulated close to the tube center due to gyrotaxis drive flow faster there and cause the vorticity profile to deviate from zero around the origin more strongly than for the solutions with $\Ri=0$ (Figure \ref{fig:cellconc}c). The larger the value of $n_{av}$, the sharper the accumulation and the stronger the flow deviation, $\chi$. This does not occur in model F because cells on average spend more time by the walls (in the radial cell distribution tails). This prevents accumulations that would lead to significant reciprocal coupling via negative buoyancy. The model F distributions for concentrated cell suspensions (with nonzero reciprocal coupling) differ little from the $\Ri=0$ case (Figure\ref{fig:cellconc}b and c, inset).

\begin{figure}
\adjincludegraphics[height=0.42\linewidth,trim={0 0 12mm 0},clip]{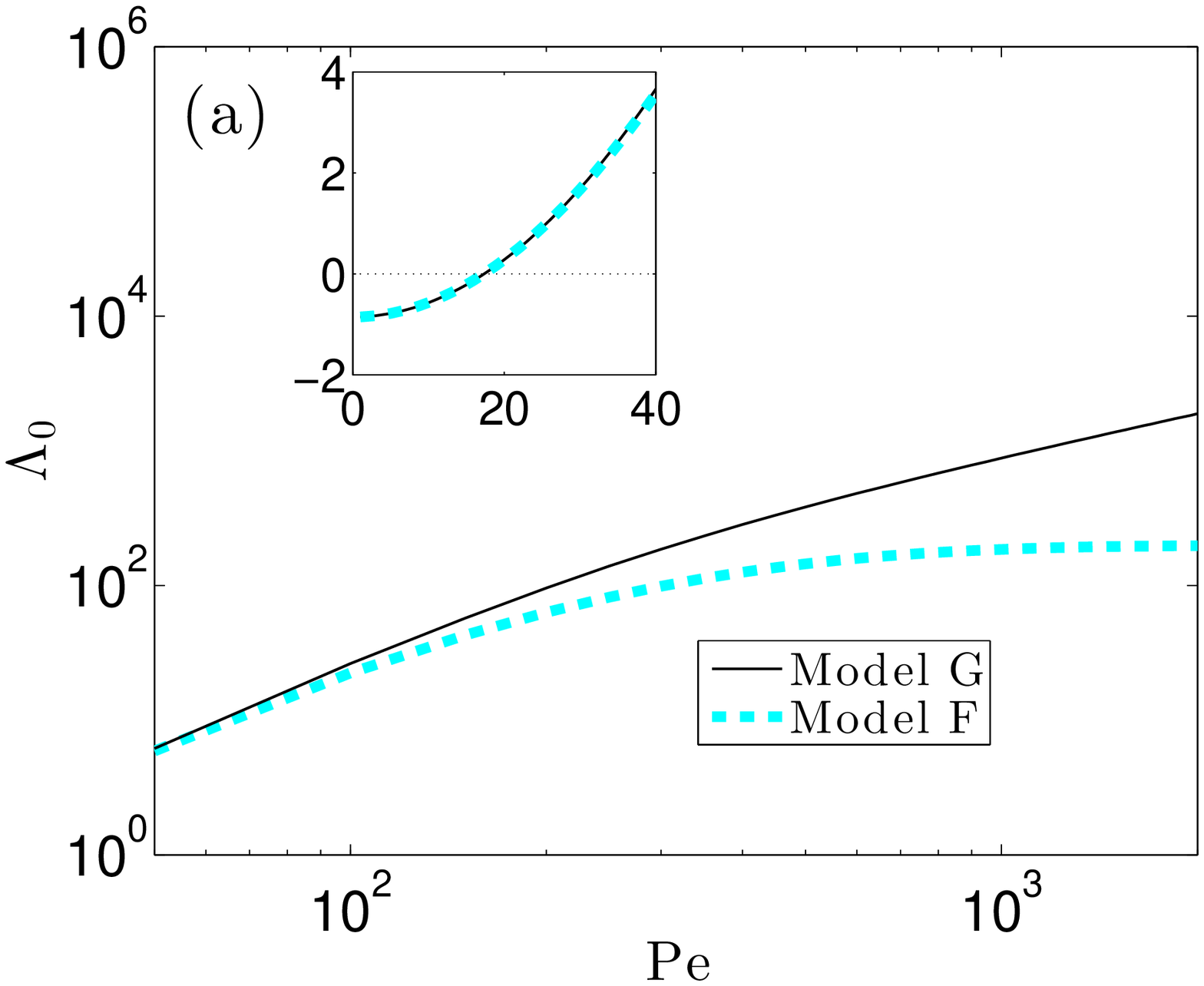}
\includegraphics[width=0.5\linewidth]{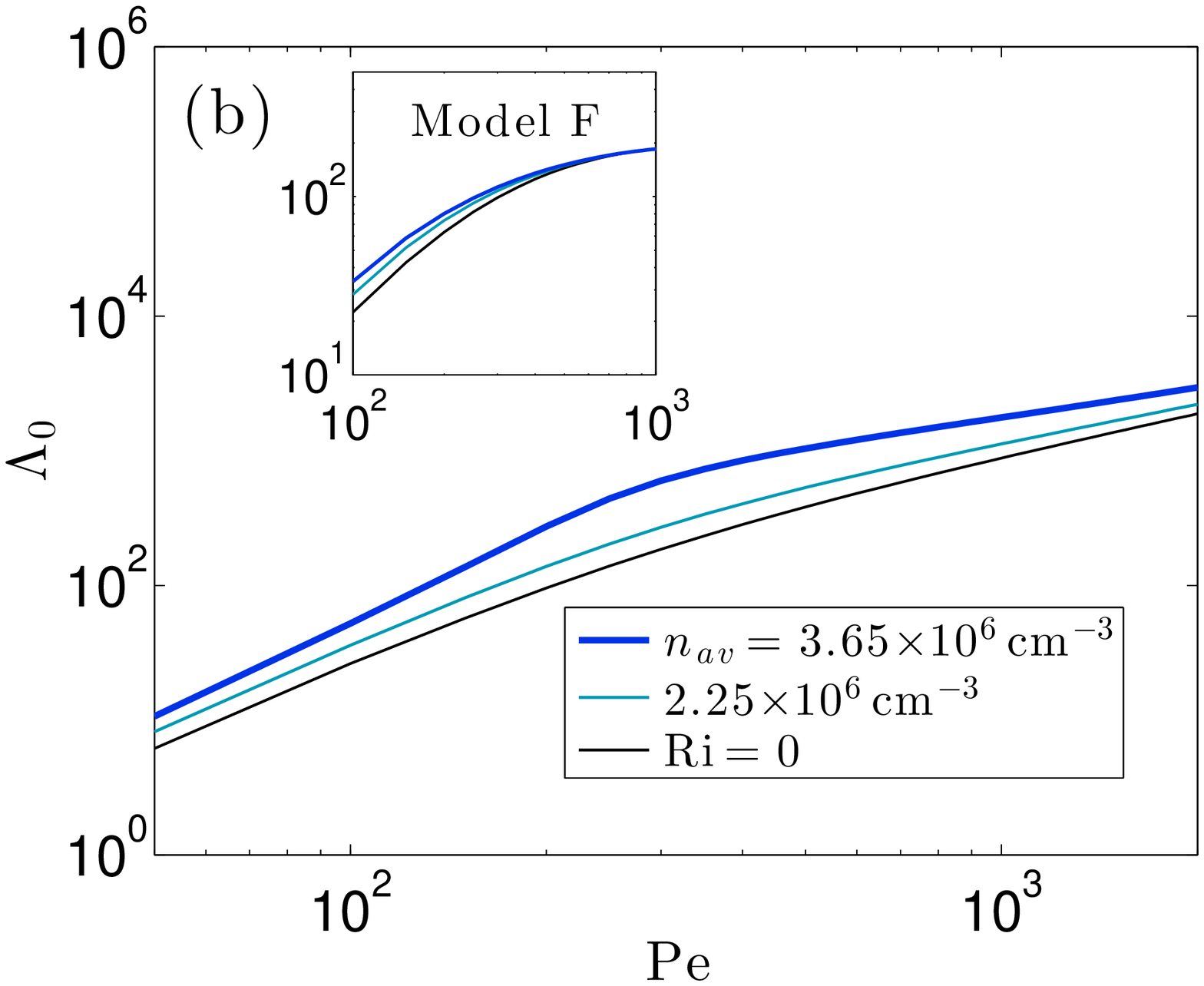}\\
\includegraphics[width=0.5\linewidth]{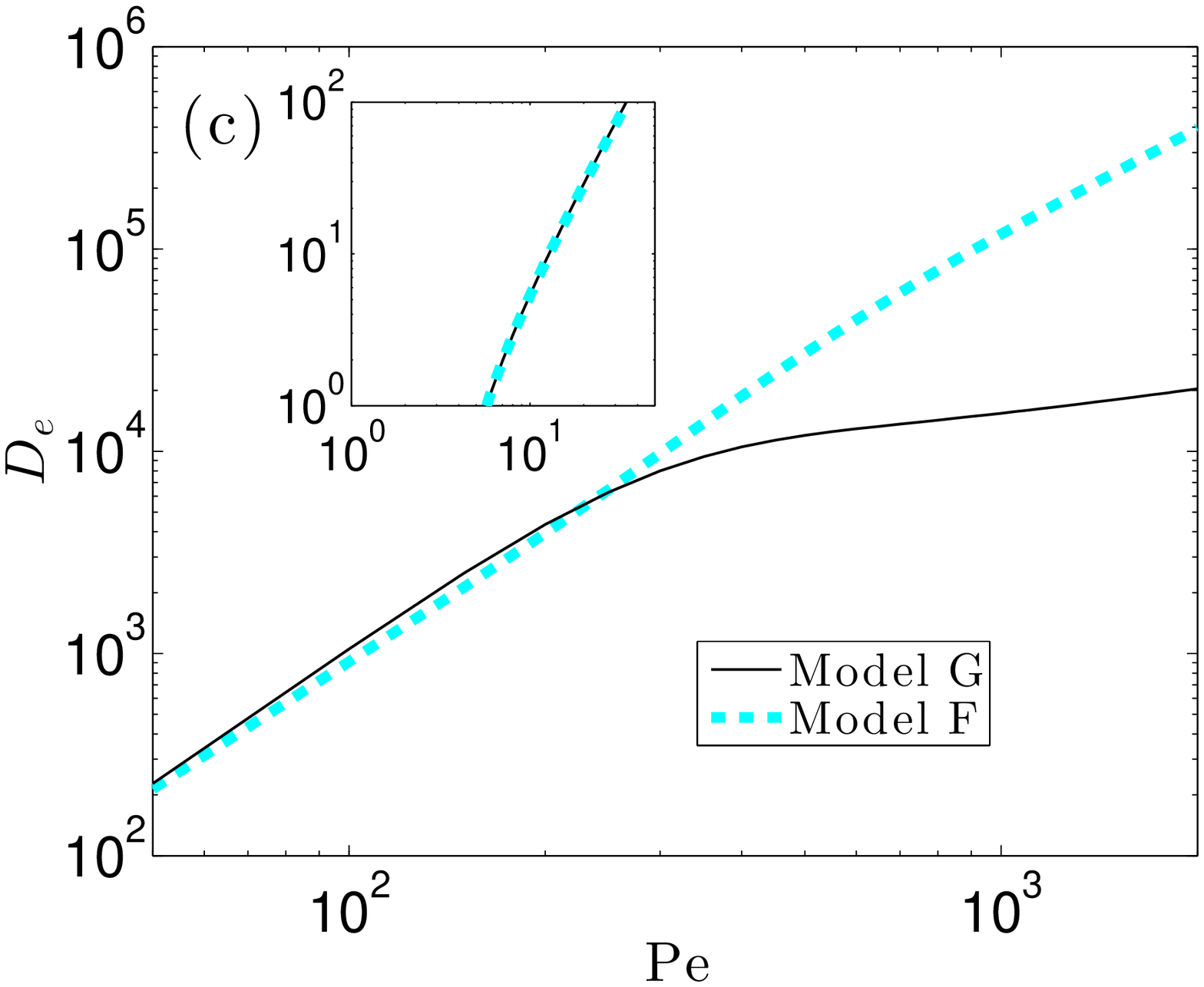}
\includegraphics[width=0.5\linewidth]{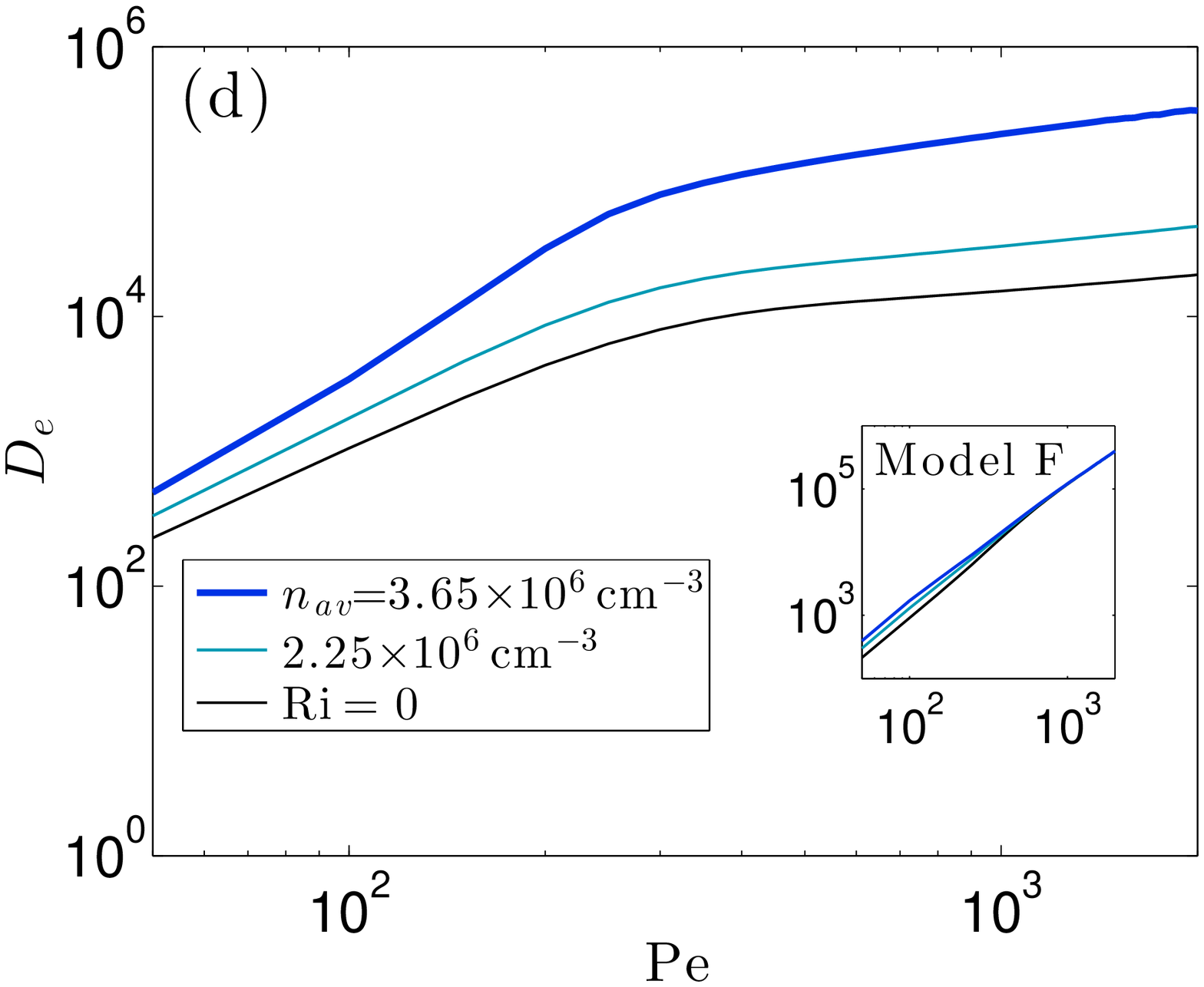}
\caption{Theoretical dispersion measures: drift and diffusivity in the absence (a and c) and presence (b and d) of reciprocal coupling via buoyancy effects. Models F (dashed) and G (solid) agree at low values of Pe for neutrally buoyant cells (a, c insets, where overlapping F and G predictions are shown), but predict different drift behaviour for large Pe. Coupling via negative cell buoyancy quantitatively changes the predictions of model G: negative buoyancy increases cell drift and enhances diffusivity (greater differential advection due to modified vorticity profile, see text) with respect to the neutrally buoyant case. On the other hand, coupling affects model F predictions very little (b, d insets).}\label{fig:dispersion}
\end{figure}

The distribution of flow and cells determines dispersion, measured by the steady state drift above the mean flow and effective diffusivity; the skewness vanishes at long times \citep{BeesCroze10}. Figure \ref{fig:dispersion}a presents a comparison of the  dimensionless drift above the mean flow, $\Lambda_0$, predicted for models F and G in the absence of reciprocal cell-flow coupling. Both models predict $\Lambda_0>0$ for sufficiently large \Pe, the distinguishing characteristic of gyrotactic dispersion \citep{BeesCroze10, BearonBeesCroze12, Crozeetal13}. As expected, at low Pe (\ref{fig:dispersion}a, inset) the F and G predictions coincide and, for very low Pe, $\Lambda_0$ is negative because cells are oriented upwards \citep{BeesCroze10}. At higher Pe, the predictions of the two models diverge: in G, $\Lambda_0\sim\rm{Pe}$; in F, it tends to a constant. The reason for this is evident from the distributions of Figure \ref{fig:cellconc}a. The increased concentration around the centreline with Pe in model G causes the larger drift (more cells in fast flow). In model F, on the other hand, the number of cells in the fast central zone of the pipe saturates, and with it the drift. Reciprocal cell-flow coupling enhances the focusing of the distribution for model G, as well as the magnitude of the flow, so a greater value of $\Lambda_0$ is possible for the same Pe. In model F, coupling makes little difference to the drift (Figure \ref{fig:dispersion}b, inset).

The predicted effective diffusivity of the dispersing population is shown in Figure \ref{fig:dispersion}c in the absence of reciprocal coupling. It increases monotonically in this Pe range for both the F and G models, which, as previously, agree for very low Pe. However, we see crucial differences. In model G, $D_e$ appears to saturate, while in model F it continues increasing with Pe. To understand this, note that the axial dispersion measured by $D_e$ is caused principally by diffusion across streamlines and differential advection of cells distributed by gyrotaxis. In model G, as Pe is increased, the cell distribution sharpens and less cells are exposed to the high shear close to the pipe wall; the increase of $D_e$ thus drops at high Pe. In model F, on the other hand, the distribution tails ensure cells will reside in regions of high shear, so $D_e$ increases with Pe. The dispersion in the presence of coupling is shown in  Figure \ref{fig:dispersion}d. We see that in model G, larger cell concentrations provide a larger $D_e$ for the same Pe. The greater vorticity in the flow close to the tube center (providing greater differential advection) contributes to a greater dispersion in the reciprocal coupling case. 
The effect of reciprocal coupling on diffusion is not very significant in model F, whose predictions for $D_e$ are not very different from the uncoupled model (Figure \ref{fig:dispersion}d, inset), in contrast to model G.

%

\subsection{Connection with experiment \label{sec:exp}}

Experimentally we measure the absolute drift of a slug of fluorescently labelled cells within a plume, given by $V_d=\frac{d Z}{dt}$, 
where $Z$ is the vertical position of the peak intensity of recorded plume (see experimental methods). A useful measure of the drift is the fractional drift above the mean flow, $\delta = \frac{V_d - U}{U}$, where $U$ is the mean flow speed. This fractional drift can be compared with the prediction from the dispersion model. Recall that the theoretical drift is defined in (\ref{driftDetrans}) as $\Lambda_0=\frac{d  \hat{m}_1}{d \hat{t}}$, with $\hat{t}=t (a^2 D_r/V_s^2)^{-1}$, $\hat{m}_1=\frac{\langle Z \rangle -U t}{a}$, and Pe defined in (\ref{eq:Pe}). It then follows that the theoretical fractional drift is simply given by $\delta = \frac{\Lambda_0}{\rm{Pe}}$.

\section{Experiments}\label{sec:experiments}
\subsection{Methods}\label{sec:methods}

{\it Algae and culture}. Batch cultures of {\it Dunaliella salina} (CCAP 19/18) were grown on a $12$:$12$ light/dark cycle on a modification of the medium of \cite{Picketal86}, which we will denote herein as DSM ({\it Dunaliella salina} medium). Healthy motile cells were gravitactically concentrated overnight on rafts of cotton wool at the surface of a culture, in a similar fashion to \cite{CrozeAshrafBees10}. These cultures were kept under the same light cycle as batch cultures to preserve circadian rhythms. Light intensity, however, was reduced to minimise the disruption of upswimming by phototaxis during concentration. Cells were harvested using a Pasteur pipette. Cell concentrations were obtained from readings of $A$, the suspension absorbance at $590$ nm, measured using a spectrophotometer (WPA CO7500), calibrated to haemocytometer counts; concentrations are given by $c=5\times 10^6\,A$ cells/cm$^3$, with fractional error $\Delta c/c=0.07$.

{\it Fluorescent staining}. Cells were stained with fluorescein diacetate (FDA, Sigma), a vital dye. We mixed 8 $\mu$l of dye from a 5 $\mu$M  stock (acetone) into 20 ml of cell suspension with absorbance typically in the range $A=0.7$-$1.5$. After letting the dye act for about $5$ min, dyed cells were `washed' by letting them swim up through a raft of cotton overlaid by a fresh layer of DSM medium. This step ensures the dye is only inside the cells. However, since fluorescein cleaved from FDA inside the cell slowly leaks out, dispersion experiments were performed within $20$ min of washing to allow macroscopic imaging with good contrast between dyed cells and the background medium.

\begin{figure}
\centerline{\includegraphics[width=0.6\linewidth, angle =-90]{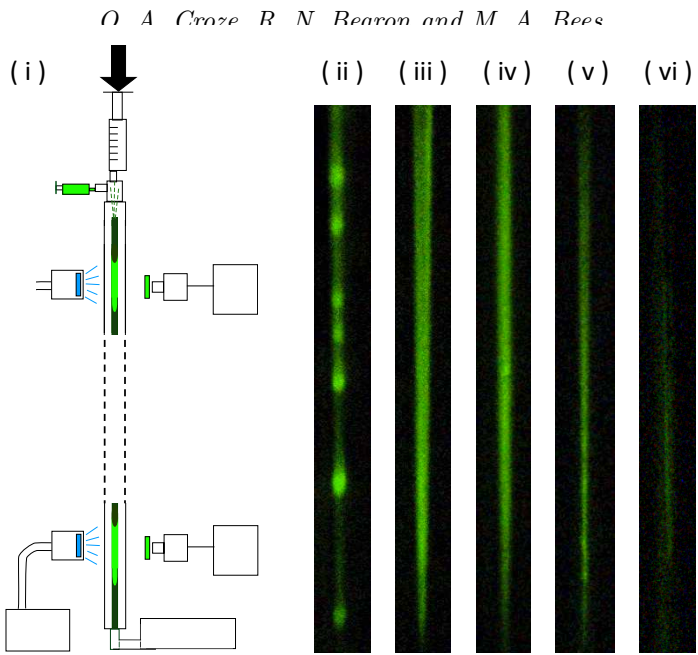}}
\caption{(i) Experimental set-up: fluorescently labelled slugs of {\it D. salina} dispersing in pipe flow, driven by a syringe pump. Labelled cells are injected near the top of the tube, recruited to an existing plume, and imaged downstream. Image (ii) shows a blippy slug for a flow rate $\Phi=100$ml  h$^{-1}$, $n_{av}=3.65 \times 10^6$ cells/cm$^3$ at $L=$60 cm from injection $[\Phi, n_{av}\,, L]=[100, 3.6, 60]$. The next images show blip-less plumes: (iii)  [$200$, $4.5$, $60$], (iv)  [$200$, $3.65$, $100$], (v) [$300$, $2.95$, $100$], and (vi) [$400$, $2.95$, $100$] (faint). Images are $1.1$ cm $\times 11.0$ cm.}\label{fig:setup}
\end{figure}

{\it Experimental set-up and imaging}. A schematic of the apparatus is shown in figure \ref{fig:setup}i. A Graseby 3500 syringe pump (Graseby Medical Ltd., Watford, UK) attached at the top of the set-up drives flow in a vertically oriented perspex tube that is $2$ m in length and $7$ mm in diameter. Experiments were performed with flow rates in the range $100-400$ ml h$^{-1}$ at a stable ambient temperature of $T=24\pm1^{\circ}$C. Sub-populations of {\it D. salina}, fluorescently labelled as above, were introduced into the flow using a $1$ ml syringe at the top of the tube, $2$ cm downstream from the pump; dyed cells were injected into a steady flow of unlabelled cells. The ensuing dispersion does not depend on the rate of injection because this rate, $\sim 10^3$ ml h$^{-1}$, was much larger than the largest advection flow rate used. The absorbance of labeled and unlabelled populations was matched. 
We imaged the ensuing dispersion using a GE680C CCD camera (Prosilica) and excitation/emission filters (MF475-35/MF525-39, Thorlabs) suitable for the fluorescence of fluorescein from FDA whose absorption/emission peaks are $495$ nm/$520$ nm (blue/green). The imaging was centred on a point $L=60$ or $100$ cm below the inoculation point. Sequences were captured at $2.1$ Hz. 

{\it Image analysis}. Dispersion measures were obtained from captured image sequences of descending slugs. The total pixel intensity in four regions of interest (ROI) of an image was calculated for each image in a sequence of a given plume. The time $t$ at which the intensity peaked corresponded to the time at which the plumed passed the vertical position $Z$ (on which the ROI was centred). Thus, a linear fit to a plot of $Z$ versus $t$ provides a value for the plume drift speed $V_d=d Z/dt$. As discussed in the next section, it was very challenging to obtain non-blippy, high contrast plumes. 
For low contrast plumes, it was possible to estimate the drift from the displacement of the plume endpoints. The error in the drift was estimated as the standard deviation over measurements with the same sample (same mean concentration $n_{av}$).

\subsection{Plumes and blips}\label{sec:blips}

In order for the experiments to test meaningfully the theory of gyrotactic dispersion, the fluorescently stained cells have to be recruited to the plumes formed by non-stained cells. It is clear from the images shown in Figure \ref{fig:setup}ii-vi that fluorescent {\it D. salina} swimmers are indeed recruited to the plumes. A blip instability has occurred in Figure (\ref{fig:setup}ii): fluorescence is localised in the blip nodules.
The theory of dispersion described in section \ref{sec:theory} assumes axial invariance of the background cell concentration and flow, which is broken when blip instabilities arise. The instability occupies certain regions of the flow-concentration parameter space, as predicted for channel flows for {\it C. augustae} parameters by \cite{HwangPedleyBlips14}. A similar stability analysis has not been carried out for {\it D. salina} in pipe flow. The dispersion experiments are challenging because of competing constraints. In particular, moderately concentrated suspensions are necessary for good fluorescent image contrast, but high concentrations are more likely to produce blips (by increasing the suspension Richardson number). However, with some care, it is possible to find blip-free plumes over the length and time scales of dispersion experiments. 

\subsection{Results: drift above the mean flow}

\begin{figure}
\centerline{\includegraphics[width=0.7\linewidth]{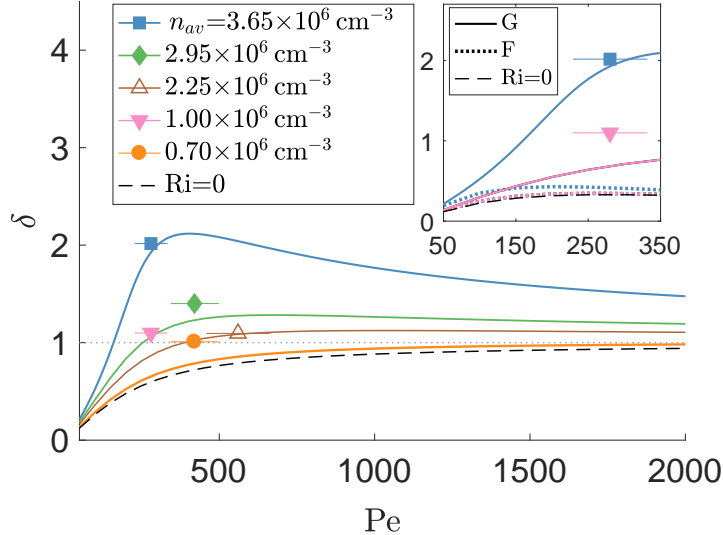}}
\caption{Comparison of theoretical predictions and preliminary experimental observations of fractional drift in dispersing slugs of {\it D. salina}, for different background cell concentrations, $n_{av}$. Same colour lines (theory) and symbols (experiment) denote equal concentrations. Quantitatively, predictions using microscopic model G (Generalised Taylor dispersion) {\it and} accounting for cell-flow coupling agree much better with experiment than using model F (Fokker-Planck model), even for the lowest Pe data (inset, symbols/line-types as in main figure). Dashed black lines indicate predictions neglecting reciprocal coupling due to negative buoyancy ($\rm{Ri}=0$). For clarity, the model G prediction for $n_{av}=10^6$ cells cm$^{-3}$ (pink line) is shown only in the inset. Vertical error bars (from same-run repeats) are smaller than the data point symbol size. Horizontal error bars stem from uncertainties in $d_r$ (see Appendix \ref{sec:appCmicro}).}\label{fig:drift_thvsexp}
\end{figure}

The dispersion data were used to evaluate the drift speed, and hence the fractional drift above the mean flow, $\delta=(V_d - U)/U$, where $V_d$ is the measured drift and $U$ is the mean flow speed, as described in the methods. The predictions from the reciprocally coupled dispersion theory for the F and G models are shown in Figure \ref{fig:drift_thvsexp} for different values of the background mean cell concentration, $n_{av}$.
In the absence of reciprocal cell-flow coupling ($\rm{Ri}=0$), the model G fractional drift $\delta$ increases monotonically with Pe, increasing to $1$ for large Pe. The model F prediction on the other hand at first increases before reaching a peak and decreasing to zero (see Figure \ref{fig:driftmodelF} in Appendix \ref{sec:appFdriftdata}). We can interpret these results in terms of the distribution of cells, in a similar manner to how we interpreted the drift in Figure \ref{fig:dispersion}a. The cells in Model G can be sharply focused to travel close to the maximum flow speed $V_d=2U$ ($\delta=1$). In model F, on the other hand, $\delta$ tends to zero at large Pe (see Figure \ref{fig:driftmodelF} in Appendix \ref{sec:appFdriftdata}), reflecting the uniformity of the cell distribution. Coupling for model G increases the value of $\delta$ significantly above $1$ at intermediate Pe, but, since $\rm{Ri}\sim 1/\rm{Pe}$, at large Pe the fractional drift tends to $1$. For model F, reciprocal coupling has little effect due to the uniformity of the cell distribution. 

The experimental results provide good evidence for $\delta>0$: the dispersion of gyrotactic swimmers is not the same as that of passive tracers. Furthermore, the experimental drift values are consistent with the predictions of model G in the presence of reciprocal cell-flow coupling. 
Model F (with or without coupling) agrees poorly with the experimental observations, even for the lowest experimental values of Pe studied (Figure \ref{fig:drift_thvsexp}, inset). The data suggest that it is necessary to include reciprocal cell-flow coupling in theoretical descriptions of fast flowing suspensions of gyrotactic swimmers. 



\section{Discussion and outlook}\label{sec:discussion}

We have carried out the first experimental test of the theory of gyrotactic dispersion in pipe flow and associated stochastic microscopic models.  We injected {\it Dunaliella salina} algae, fluorescently labelled by fluorescein diacetate, into an unlabelled population in downwelling pipe flow, and imaged the dispersing `slugs' to measure their fractional drift above the mean flow. Experimental drift values were compared with predictions from an axial dispersion model \citep{BeesCroze10, BearonBeesCroze12} with swimming and buoyancy parameters measured using tracking microscopy. The dispersion model requires as inputs transport functions (local mean orientation and cell diffusivity tensor) dependent on the local shear rate, which can be evaluated from stochastic microscopic models. Predictions using two such models, referred to as model F \citep{PedleyKessler90, PedleyKessler92} and G \citep{HillBees02, ManelaFrankel03, BearonBeesCroze12}, were employed with parameters for {\it D. salina}. We derived new solutions of the axial dispersion model combining the effects of both negative buoyancy and nonlinear solutions of the two alternative stochastic microscopic models. \citet{BeesCroze10} had previously considered only linearised transport functions for the F model with negative buoyancy.

The experimental results provide clear evidence for the nonzero drift above the mean flow expected in the dispersion of gyrtotactic swimmers \citep{BeesCroze10, BearonBeesCroze12, Crozeetal13} (see Figure \ref{fig:drift_thvsexp}). This is in stark contrast to the zero drift of passive, neutrally buoyant tracers \citep{Taylor53, Aris56}. 
Furthermore, in the Pe range studied, we find quantitative agreement between experiment and theoretical predictions obtained using model G and accounting for negative cell buoyancy (Ri$>0$). As shown in Table \ref{DriftValues} of Appendix \ref{sec:appFdriftdata}, 
three out of the five experimental points agree well with theory within the stated errors. For the other two points (corresponding to the lowest cell concentrations studied), theory underestimates the drift.
Agreement with model F, on the other hand, is poor with or without buoyancy, even for the lowest values of Pe (Figure \ref{fig:drift_thvsexp}, inset). This should be expected for the relatively large Pe used (recall model F and model G agree only at low Pe), as local spatial stochastic effects in strong shear flow, neglected in model F, become important. The superior performance of model G over model F expected at large Pe has been shown previously by \citet{Crozeetal13} who compared analytical dispersion predictions for models F and G with Lagrangian simulations. This work, however, ignored the effect of negative cell buoyancy. Without the latter, agreement between the experimental results and model G is not very good. This highlights the important role of reciprocal flow-cell coupling in the dispersion dynamics.

Given the reciprocal coupling between cells and flow it is very difficult to design experimental apparatus to test microscopic stochastic models of macroscopic suspensions of biased swimming cells.  In this manuscript we have presented one method that has shown great promise, demonstrating the shortcomings of the F model whilst providing evidence in support of the G model with the inclusion of negative buoyancy effects. However, there were some significant experimental challenges.
The experiments were limited by the constraints of good image contrast, steady dispersion and the emergence of blip instabilities. The latter proved antithetic to the former two, as high concentrations provided better contrast but also a greater propensity for blip formation for the range of Pe investigated. In addition, the use of long tubes to guarantee steady dispersion made blips more likely (giving slow growing unstable modes time to develop). Blips are known to depend sensitively on concentration and flow rate \citep{Kessler86, DennisenkoLukashuck07}, but we lack a quantitative guide to avoid the instability. The work of \citet{HwangPedleyBlips14} provides only a qualitative guide, as this study focused on {\it C. augustae}, which has different gyrotactic parameters \citep{HillHaeder97}.
 
Future experiments should refine the drift measurements and test other dispersion predictions, such as dependence of the effective diffusivity $D_e$ of the swimming algae on key flow and swimming parameters. Fluorescent imaging with at least two downstream cameras will allow measurements of the axial variance $\Var{(Z)}$ of a dispersing slug. Theory predicts $\Var{(Z)}=2 D_e V_s^2 d_r^{-1} t$ from which $D_e$ can be evaluated. The practical issues associated with fluorescent dye diffusion and loss of contrast may be circumvented using mutants (e.g. {\it Chlamydomonas reinhardtii}, expressing fluorescent protein). 
Our experience in experimenting with such approaches provides two important caveats: mutants from `normally swimming' wild-type background strains should be employed; motility should not be compromised by the insertion of fluorescent gene constructs.   

The dispersion models we have tested experimentally in this work are predicated on swimmer suspension theories developed in the early 90s (see \citet{PedleyKessler92} for a review). While several improvements have been made to these theories, they have not been subject to extensive experimental tests since their formulation. In particular, basic assumptions of the theories, such as the dilute nature of the swimmer suspensions have not been challenged experimentally. In this work, we follow the traditional assumption that swimmer suspensions are dilute, but, considering the formation of plumes, one can question if the dilute assumption is appropriate for the description of self-focusing suspensions of gyrotactic swimmers and axial dispersion.  The cell concentration at the center of the pipe depends on the applied flow rate. From Figure~\ref{fig:cellconc}b, we find that for Pe=1000 (and the parameters in Table \ref{ModPar}), the concentration at the pipe center is approximately 100 times that of the average, i.e.~$10^8$ cells/cm$^3$. This corresponds to a volume fraction of 0.02 and a mean distance between cells of $20$ $\mu$m (two cell diameters). Therefore, we might expect cell-cell interactions between cells to play a role \citep{Drescheretal11}. However, we note that the cell concentration drops very rapidly from the center, so high concentration effects should match the dilute theory except for a narrow core at the center of the pipe, which is advected at close to the maximum flow rate. Furthermore, axial cell diffusivity is dominated by cell motion in the high shear regions and not within the narrow core.  Consequently, the axial dispersion predictions of the dilute theory are likely to hold even for sharply focused plumes.  The rationale of this work was precisely to ascertain how well the predictions from dilute theory perform by testing them experimentally.  The experimental results suggest that the dilute assumption is not  major issue.  However, it would be worthwhile to extend the theory to explore a semi-dilute regime.

Another assumption made in the original theories (again, refer to \citet{PedleyKessler92}) is that the swimmer population heterogeneity is unimportant. It is assumed that averages of swimming parameters over a population (e.g. mean swimming speed) are adequate to describe the suspension behaviour and fluid mechanics. This will clearly only be a good approximation if parameter distributions are sufficiently narrow. \citet{Beesetal98} studied theoretically the effect of deviations from narrow distributions on gyrotactic transport in the F model. They evaluated how the ratio of speed variance to mean, $\langle V^2\rangle/\langle V\rangle^2$ ($V$ denotes speed, angled brackets an average over the population), changes the components of the diffusivity tensor $\mathbf{D}$. We note that, as for the dilute suspension assumption, the agreement we find between theory and experiments seems to indicate that the neglect of swimming population heterogeneity is not a serious one in the dispersion of {\it D. salina}. This cannot, however, be generalised to other swimming species, which may have broad swimming parameter distributions. Furthermore, even within a particular species, biological changes, such as metabolism, can drive significant heterogeneity. It would be of interest to consider how swimming parameter distributions and their changes affect suspension dynamics and dispersion.

Accurately predicting the unusual dispersion of swimming algae in pipe flow is important for the design of more efficient and `considerate' photobioreactors for their cultivation \citep{BeesCroze14}. Closed photobioreactor designs comprise channels or pipes in which growing suspensions of algae are flowed. In tubular external loop air-lifts, cells are bubbled with air in a riser tube and then recirculated in a downcomer connected to the riser by two horizontal tubes. For gyrotactic biotechnological species of interest, such as the $\beta$-carotene producer {\it D. salina} considered here, the dispersion observed in this work is very relevant. For example, the non-zero drift above the mean flow we have observed will cause nutrients or waste (passive tracers) to become separated from the cells in the downcomer. Experiments are in progress quantifying the behaviour of gyrotactic swimmers in lab and pilot-scale photobioreactors. More complex dispersion may ensue in photobioreactors operated in turbulent pipe flow. Simulations predict gyrotactic dispersion due to cells focusing transiently in downwelling regions of the turbulent channel flows \citep{Crozeetal13}, but experiments on swimmers in bounded turbulent flows have not yet been carried out. It will be of considerable ecological interest to explore the consequences of gyrotactic dispersion in river, lake and oceanic flows. 

\section{Acknowledgements}

We thank Peter Dominy and John Christie for advice on fluorescent vital dyes; Graham Gibson for help with fluorescent imaging; and Jochen Arlt and Dario Dell'Arciprete for help with videomicroscopy of {\it D. salina}. We also acknowledge discussions with Pietro Cicuta, Silvano Furlan, Yongyun Hwang, Di Jin, Vincent Martinez, Tim Pedley and Emma Silvester. OAC and MAB gratefully acknowledge support from the Carnegie Trust for the Universities of Scotland and EPSRC (EP/D073398/1). OAC acknowledges support from a Royal Society Research Grant and the Winton Programme for the Physics of Sustainability.

\appendix

\section{Stochastic models and shear-dependent transport} \label{sec:appAqrDrr}

We summarise very briefly the mathematical structure of models F (the orientation-only Fokker-Planck model)  \citep{PedleyKessler90, PedleyKessler92} and G (Generalised Taylor dispersion) \citep{HillBees02, ManelaFrankel03}. The reader is referred to the original literature for more details.  

Consider $P(\mathbf{p},\mathbf{x},t)$, the probability of finding a cell with orientation $\mathbf{p}$ at position $\mathbf{x}$ at time $t$. This evolves according to \citep{HillBees02}
\begin{eqnarray}
\label{eq:microscale}
\frac{\partial P}{\partial t} 
+\nabla_\mathbf{x}\cdot[(\mathbf{u}+V_s\mathbf{p})P]+d_r\nabla_\mathbf{p}\cdot[
\dot{\mathbf{p}}P-\nabla_\mathbf{p} P]=0,
\end{eqnarray}
where $\mathbf{u}$ is the fluid velocity, $V_s$ is the constant cell swimming speed, $d_r$ is the rotational diffusivity due to the intrinsic randomness in cell swimming, and $\nabla_\mathbf{x}$ and $\nabla_\mathbf{p}$ are physical and orientational gradients, respectively. For gyrotactic cells
\be
\dot{\mathbf{p}}=\lambda(\mathbf{k}-(\mathbf{k}.\mathbf{p})\mathbf{p})-\sigma\mathbf{j} \wedge \mathbf{p} \label{eq:pdot}
\ee
where, as in the main text, $\lambda=\frac{1}{2d_rB}$ is the stochasticity parameter and $\sigma=-\chi^\prime\Pe/(2 \beta^2)$ is the dimensionless shear strength. The unit vectors $\mathbf{k}$ and $\mathbf{j}$ are in the upwards (antiparallel to gravity) and positive voriticity direction, respectively.  
 
\subsection{Model G}
\label{sec:GTD}

By calculating the moments of the distribution function $P$, \citet{FrankelBrenner91,FrankelBrenner93} showed that, on timescales long compared to $1/d_r$, the concentration of cells, $n(\mathbf{x},t)$, in a homogeneous shear flow satisfies an advection-diffusion equation \citep{HillBees02}: 
\begin{eqnarray}
\label{eq:ad_diff_in_flow}
\frac{\partial n}{\partial t}+\nabla_\mathbf{x}\cdot\left[\left(\mathbf{u}+V_s\mathbf{q}\right)n
-D_c\mathbf{D}\cdot\nabla_\mathbf{x}n\right]=0,
\end{eqnarray}
where $D_c=V_s^2/d_r$ is the magnitude of the swimming diffusivity. The mean swimming direction, $\mathbf{q}$, and non-dimensional diffusivity tensor, $\mathbf{D}$, can be written as integrals over cell orientation, $\mathbf{p}$ \citep{HillBees02,ManelaFrankel03}, such that
\begin{eqnarray}
 \label{eq:def_p}
\mathbf{q}&=&\int_\mathbf{p}\mathbf{p}  f(\mathbf{p} ) d\mathbf{p} ,\\
\label{eq:pos_def_diffusion}
\mathbf{D}&=&\int_\mathbf{p} [\mathbf{b}\mathbf{p}+\frac{2\sigma}{ f(\mathbf{p})}\mathbf{b}\mathbf{b}.\hat{\mathbf{G}}]^{sym}d\mathbf{p}.
\end{eqnarray}
Here $[]^{sym}$ denotes the symmetric part of the tensor, and $\hat{\mathbf{G}}$ denotes the non-dimensional fluid velocity gradient. The equilibrium orientation, $f(\mathbf{p})$, and vector $\mathbf{b}(\mathbf{p})$ are given by \citep{HillBees02,ManelaFrankel03}
\begin{eqnarray}
\label{eq:f_eqn_with_flow}
\nabla_\mathbf{p}\cdot[
\dot{\mathbf{p}}f-\nabla_\mathbf{p} f]&=&0,\\
\label{eq:b_eqn_with_flow}
\nabla_\mathbf{p}.(
\dot{\mathbf{p}}\mathbf{b}-\nabla_\mathbf{p} \mathbf{b})-2\sigma\mathbf{b}.\hat{\mathbf{G}}&=&f(\mathbf{p})(\mathbf{p}-\mathbf{q}),
\end{eqnarray}
subject to the integral constraints $\int_\mathbf{p} f d\mathbf{p}=1$ and $\int_\mathbf{p} \mathbf{b}d\mathbf{p}=0$. Equations (\ref{eq:f_eqn_with_flow}-\ref{eq:b_eqn_with_flow}) follow from evaluating the moments of the distribution function $P$ \citep{HillBees02,ManelaFrankel03}.

The orientation distribution $f(\mathbf{p})$ represents the steady state distribution of swimmer orientations. The meaning of the vector $\mathbf{b}(\mathbf{p})$ is less intuitive. It can be written as $\mathbf{b}=f \mathbf{B}d_r/V_s $, where $\mathbf{B}(\mathbf{p})$ quantifies the difference between the average position of a swimmer, given its instantaneous orientation is $\mathbf{p}$, and its average position averaged
over all values of $\mathbf{p}$ \citep{HillBees02}.

\subsection{Model F}
\label{sec:modelF}

In model F, the same advection-diffusion equation (\ref{eq:ad_diff_in_flow}) is assumed, but the transport functions are obtained from a PDF in terms of orientation only. The distribution $f$ for the orientation is governed by (\ref{eq:f_eqn_with_flow}) with (\ref{eq:pdot}). The solution for the mean orientation (which only depends on $f$), is the same as in model G. However, in model F the diffusivity tensor is given by the approximation
\begin{eqnarray}
\mathbf{D}_{F}=D_c \int_\mathbf{p}(\mathbf{p}-\mathbf{q})^2  f(\mathbf{p} ) d\mathbf{p},
\end{eqnarray}
where now $D_c=V_s^2\tau$, and $\tau$ is a direction correlation time estimated from experiment. This is not as straightforward as measuring the rotational diffusivity $d_r$ required for the G model. However, following \citet{BearonBeesCroze12}, we can use the fact that model F and G must coincide asymptotically at low shear, so that $\tau= J_1 / ( K_1 \lambda d_r )$, where values of $\lambda$ and $d_r$ can be found from experiment (see Appendix \ref{sec:appCmicro}), and $J_1(\lambda)$ and $K_1(\lambda)$ are rationally derived functions that can be calculated (for more details, please refer to Appendix D of \citet{BearonBeesCroze12}; \citet{PedleyKessler90}).

\subsection{Model solutions}
\label{sec:solutions}

The dispersion model requires knowledge of the functional dependence on dimensionless shear $\sigma$ of the components of the transport tensors $\mathbf{q}$ and $\mathbf{D}$. We follow the approach of \citet{BearonBeesCroze12} where these components are evaluated from solutions of models F and G using a Galerkin method. The solutions depend on the bias parameter $\lambda=1/(2 d_r B)$ evaluated from tracking experiments. Previously, solutions were obtained for models parametrised for {\it C. augustae} algae. Here we present new results for {\it D. salina}, for which $\lambda=0.21$ (see Table \ref{ModPar} in main text). For analytical convenience, these solutions are then fitted with rational functions.  The relevant components for dispersion are: $q^r(\sigma)=-\sigma P(\sigma; {\bf a}^r,{\bf b}^r)$, $q^z(\sigma)= -P(\sigma; {\bf a}^z,{\bf b}^z)$; ; $D^{rr}_m(\sigma)= P(\sigma; {\bf a}^{rr},{\bf b}^{rr})$; $D^{zz}_m(\sigma)= P(\sigma; {\bf a}^{zz},{\bf b}^{zz})$;$D^{rz}_m(\sigma)= -\sigma P(\sigma; {\bf a}^{rz},{\bf b}^{rz})$, where the rational function $P(\sigma; {\bf a},{\bf b})$ is given by
 \begin{eqnarray}
\label{eq:P_rat_func}
P(\sigma; \mathbf{a},\mathbf{b})=\frac{a_0 +a_2 \sigma ^2+a_4\sigma^4}{1+b_2 \sigma^2+b_4 \sigma^4}.
\end{eqnarray}
The parameters $a_0$, $a_2$, $a_4$, $b_2$ and $b_4$, are chosen to ensure the polynomials match asymptotic solutions to the stochastic equations of model F and G \citep{BearonBeesCroze12}.

\begin{table}
\begin{center}
\def~{\hphantom{0}}
\begin{tabular}{l||c|c|c|c|c}
 & $a_{0}$ &$a_{2}$ & $a_{4}$&$b_{2}$ & $b_{4}$\\
\hline
$\mathbf{a}^r$	&$3.48 \times 10^{-2}$    &$4.62 \times 10^{-3}$	&$0 $ &$ 3.80 \times 10^{-1}$	&$3.30 \times 10^{-2}$\\
$\mathbf{a}^z$		& $6.98 \times 10^{-2}$ 	&$7.71 \times 10^{-3}$	&$0$			&$3.58  \times10 ^{-1}$	&$ 2.75 \times 10^{-2}$\\
$ \mathbf{a}^{rr}_G$	&$1.66 \times 10^{-1}$	&$-1.03 \times 10^{-4}$	&$0$ &$2.47 \times  10 ^{-1}$	&$ -1.54\times 10^{-4}$\\
$ \mathbf{a}^{rz}_G$	&$1.66 \times 10^{-1}$	& $3.37 \times 10^{-6}$	&$0$			&$4.98\times 10 ^{-1}$	&$6.19 \times 10^{-2}$\\
$ \mathbf{a}^{zz}_G$	&$1.67\times 10^{-1}$	&$3.64\times 10^{-1}$	&$1.10\times 10^{-6}$ &$4.84 \times  10 ^{-1}$	&$6.07\times 10^{-2}	 $\\
$ \mathbf{a}^{rr}_{F}$	&$1.66 \times 10^{-1}$	& $3.36 \times 10^{-2}$	&$1.15 \times 10^{-2}$&$2.04\times10 ^{-1}$		&$6.91 \times 10^{-2}$\\
$ \mathbf{a}^{rz}_{F}$	&$5.94 \times 10^{-4}$	& $0$	&$0$	&$4.34\times 10 ^{-1}$	&$2.54 \times 10^{-1}$\\
$ \mathbf{a}^{zz}_{F}$	&$1.65 \times 10^{-1}$	&$1.04 \times 10^{-1}$				&$6.91 \times 10^{-4}$			&$6.28\times 10^{-1}$		&$4.16 \times 10^{-3}$
\end{tabular}
\caption{Parameters used in the rational function fits of solutions for models F and G used in our dispersion theory predictions.} \label{abcoeff}\label{tab:abcoeff}
\end{center}
\end{table}

\section{Modified Pick medium} \label{sec:app}

The {\it D. salina} cells were grown in a modification of Pick's medium \citep{Picketal86}, with final salt concentration of $0.5$ M. The medium consists of 
(J. Polle, private communication): a $100\times$ concentrated nutrient mix ($2$M Tris-HCl, pH$=7.5$; $1$M $\mathrm{KNO}_3$; $1$M $\mathrm{MgSO}_4$; $60$ mM $\mathrm{CaCl}_2$; $20$ mM $\mathrm{KH}_2\mathrm{PO}_4$; $0.4$ mM $\mathrm{FeCl}_3$ in $4$ mM EDTA, pH$=7.5$); a $1000\times$ concentrated trace micronutrients ($150$mM $\mathrm{H}_3\mathrm{BO}_3$; $10$ mM $\mathrm{MnCl}_2$; $0.8$ mM $\mathrm{ZnCl}_2$; $0.3$ mM $\mathrm{CuCl}_2$; $2$ mM $\mathrm{Na}_2\mathrm{MoO}_4$; $2$ mM $\mathrm{NaVO}_3$; $0.2$ mM $\mathrm{CoCl}_2$); a $4\times$ concentrated salt soliution ($2$ M $\mathrm{NaCl}$); and a $20\times$ concentrated bicarbonate soliution ($0.5$ M $\mathrm{NaHCO}_3$, filter sterilised and added after autoclaving the nutrients and salt mixed together). 

\section{Model parameters from microscopy} \label{sec:appCmicro}

The swimming speed $V_s$ and bias parameter $\lambda$ were obtained by tracking video microscopy of a suspension of {\it D. salina} swimming in a vertically oriented capillary. The mean speed $V_s$ is obtained straightforwardly from the distribution of cell displacements \citep{}. The distribution of angles to the vertical was inferred from the tracking data and fitted with a Von Mises distribution, as in \citep{HillHaeder97}:
\begin{equation}\label{VonMises}
f(\theta)=\frac{\exp[\lambda\cos(\theta-\theta_0)]}{2\pi I_0(\lambda)}.
\end{equation}
%

The fit allowed us to estimate $\lambda$. Alternatively, $\lambda$ can be found from the ratio of the mean upswimming speed $V_u$ to the mean swimming speed $V_s$, since $V_u/V_s=K_1(\lambda)=\coth(\lambda)-1/\lambda$ \citep{PedleyKessler92}. The video data was captured at $100$ fps in videos lasting $40$ s. These captured {\it D. salina} cells swimming in a $400$ $\mu$m deep flat glass chamber (CM Scientific, $50$ mm $\times$ $8$ mm $\times$ $0.4$ mm). Particle tracking using the established algorithms of \citet{CrockerGrier96} was implemented via the Kilfoil lab MATLAB routines (http://people.umass.edu/kilfoil/downloads.html). This typically provided $\approx 2500$ tracks per video that were analysed to obtain the values of $V_s$ and $\lambda$ shown in table \ref{ModPar}. Values of $\lambda$ obtained from the alternative methods discussed above from the same data were averaged; values from independent data sets were also averaged. The final error in the parameter was estimated by evaluating the standard error in the mean of these independent data.   

Tracking also provides an estimate from the gravitational reorientation time $B$. 
It was estimated from the reorientation of cells immobilised by heating the suspension to $40^{\circ}$C. Tracking the reorientation of the sedimenting cells provided $B$ by comparing with the analytical solution of equation (\ref{eq:pdot}) (with $\sigma=0$) for the reorientation of $\theta$: $\theta(t)=2\arctan[\exp(-(1/2B)(t-t_0)]$. We find $B=10.5\pm0.8$s. With $B$ and $\lambda$ known, the value for the rotational diffusivity follows from the definition of $\lambda=1/(2Bd_r)$. The errors in the tracking parameters determine the error in derived quantities. In particular, by standard error propagation it can be shown that the error in Pe is dominated by the error in the rotational diffusivity. This uncertainty $\Delta \Pe/\Pe\approx\Delta d_r/d_r=0.18$ is sizeable, and is represented as horizontal error bars in Figure \ref{fig:drift_thvsexp}.

Tracking microscopy of sedimenting cells (the same data that was used to estimate $B$) also allows to quantify their negative buoyancy. The drag force on a sedimenting cell can be estimated from $F=6 \pi \mu R U_s$ (generalised Stokes law), where $R$ is the effective hydrodynamic radius of a sphere with the same viscous resistance to flow as the cell and $U_s$ is the average sedimentation speed of the cells. These were heat-immobilised as above and their settling was captured in $100$s long videos at  $2.52$ fps and tracked (as above) to provide $U_s=(1.44\pm0.37)\times 10^{-4}$cm/s. For sedimenting negatively buoyant cells, the buoyant force $\Delta\rho \,v_c\,g$, where $v_c$ is cell volume and $g$ is the gravitational acceleration, balances the viscous drag. Thus we can estimate
\be\label{eq:deltarho}
\Delta\rho\,v_c=6 \pi \frac{\mu}{g} R U_s,
\ee
if $R$ is provided. This radius can be estimated from the angularly averaged friction factor of a prolate ellipsoid, our chosen approximation for the {\it D. salina} cells. Using an electrostatic analogy it can be shown that for ellipsoids $R=2/\xi_0$  \citep{HubbardDouglas93}, where
\be
\xi_0=\int_0^\infty\frac{1}{\sqrt{(a^2+\zeta)(b^2+\zeta)(c^2+\zeta)}} d\zeta \underrel{b=c<a}{=}\frac{2 \cosh^{-1}(a/b)}{\sqrt{a^2-b^2}}.
\ee
By image analysis of micrographs, we estimate $a=(4.8\pm2.7)\times 10^{-4}$ cm and $b(=c)=(3.2\pm1.9)\times 10^{-4}$cm. With these values $\xi_0=5.4\pm1.6\times 10^{-4}$ cm$^{-1}$ and $R=(3.7\pm1.1)\times 10^{-4}$cm, so that, 
with $\mu$ and $g$ from Table \ref{ModPar}, we can use equation (\ref{eq:deltarho}) to find $\Delta\rho\,v_c=(0.92\pm0.36)\times 10^{-11}$g. This product is all that is required to parametrise the model: knowledge of the independent values of $\Delta\rho$ and $v_c$ is not necessary. Calculating the product $\Delta\rho\,v_c$ rather than its factors has the added benefit of reducing the experimental uncertainty in our estimate with respect to the one evaluated from the factors. Apart from general values for biflagellates quoted by \citet{Kessler86}, we could not find measurements of the excess density of {\it D. salina} in the literature. However, a fluorometric study of phytoplankton sinking rates included measurements of {\it Dunaliella tertiolecta} (a close relative of {\it D. salina}), providing its excess density of $\Delta\rho=0.1$ and estimated cell volume $2.7\times 10^{-10}$ cm$^3$ \citep{Eppleyetal67}. This gives $\Delta\rho\,v_c=2.7\times 10^{-11}$ g, which is of the same order as our estimate for {\it D. salina}.

\section{Model numerical solution}\label{sec:appDnumsol}

To obtain predictions from the steady state dispersion model we need to solve equations (\ref{drifteff}) and (\ref{Deff}) with the coupled flow and cell concentrations given by (\ref{eq:NSd}) and (\ref{eq:cellsd}), subject to the constraints (\ref{eq:chiR00norm}). As mentioned in the main text, this problem is more simply tackled by writing down the equivalent differential equation system (a boundary value problem). Recalling primes denote derivatives with respect to $r$, the dispersion ODE system is given by
\bea
&& \frac{d \chi^\prime}{dr} = -\frac{1}{r}\chi^\prime + P_z - \Ri R_0^0\label{eq:dispsystflow}\\
&& \frac{d R_0^0}{dr}=\beta\frac{q_r}{D^{rr}} R_0^0\label{eq:dispsystR00}\\
&& \frac{d m_0^*}{dr}=2 r R_0^0\label{eq:dispsystm0star}\\\ 
&& \frac{d X}{dr} = 2 r \chi \label{eq:dispsystchicons}\\
&& \frac{d P_z}{dr}  = 0 \label{eq:dispsystPz}\\ 
&& \frac{d \Lambda_0^*}{dr}=2 r \left[  D^{rz} R_0^{0\prime}+ \left( \Pe \chi + \beta q^{z} \right) R_0^0 \right]\label{eq:dispsystLam0star}\\
&& \frac{d g}{dr}=\beta \frac{q^r}{D^{rr}} g+ R_0^0 \frac{D^{rz}}{D^{rr}}-\frac{\Lambda_0^*-m_0^* \Lambda_0}{2 r Drr}\label{eq:dispsystg}\\ 
&& \frac{d D_e}{dr}=2 [g^\prime+(\Pe \chi+\beta q^z-\Lambda_0) g+ D^{zz} R_0^0]\label{eq:D_e},
\eea  
subject to the boundary conditions: $\chi^\prime(0) = 0; \chi(1) = -1; m^*(0)=0; m^*(1)=1; X(0)=0; X(1)=0; \Lambda_0^*(0)=0; g(0)=0; D_e(0)=0$.

Equations (\ref{eq:dispsystflow}) and (\ref{eq:dispsystR00}) correspond to equations (\ref{eq:NSd}) and (\ref{eq:cellsd}), respectively, for the coupled flow deviation above the mean $\chi$ and (normalised) cell concentration $R_0^0$. Equation (\ref{eq:dispsystm0star}) corresponds to (\ref{eq:Lam0starm0star}) for $m_0^*$, the cumulative cell concentration within a radius $r$. Equation (\ref{eq:dispsystchicons}) implements the constraint $\overline{\chi}=0$ (from the definition of $\chi$) and (\ref{eq:dispsystPz}) states no pressure gradient is imposed. Equation (\ref{eq:dispsystLam0star}) corresponds to (\ref{eq:Lam0starm0star}) defining the partial drift. The full drift is then simply evaluated as  $\Lambda_0^*(1)=\Lambda_0$. Equation (\ref{eq:dispsystg}) corresponds to equation  (\ref{eq:g}) for the diffusivity weight function $g$ and, finally, (\ref{eq:D_e}) corresponds to equation (\ref{Deff}) for the effective axial diffusivity. Because this equation requires the value of the full drift, the system   (\ref{eq:dispsystflow}-\ref{eq:dispsystLam0star}) needs to be solved once, before solving for the diffusivity.  Numerical solutions of the ODEs were obtained using the MATLAB (Mathworks, Natick, MA, USA) bvp4c routine. So that this routine could handle the singular boundary value problem in equation (\ref{eq:dispsystg}), we carried out the change of variables $\tilde{g}=g/r$ 
and $\tilde{D_e}=D_e/r$. 
Solutions to the model equations were found using both model F and G solutions for $q^r(\sigma)$ and $D^{rr}(\sigma)$, see Appendix \ref{sec:appAqrDrr}.

\section{Blip instability}\label{sec:appEblips}

\begin{figure}
\centerline{\includegraphics[width=0.5\linewidth]{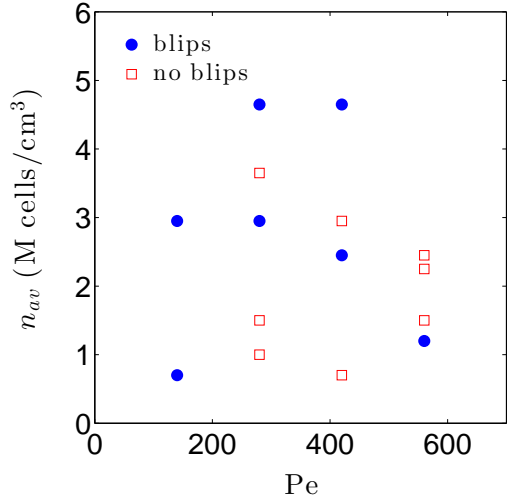}}
\caption{Chart recording the concentration-flow values giving blips or clean plumes at the time of observation of the dyed slug. This should not be considered a stability diagram: the syringe pump flow rate did not allow exploration of slow growing instabilities.}\label{fig:blipdiag}
\end{figure}

As indicated in the main text, several competing constraints made the observation of high-contrast non-blipping plumes very challenging. Figure \ref{fig:blipdiag} shows a diagram charting the parameter values where blips were observed. Note that sometimes blips were simply observed and image sequences not recorded for analysis (either for practical reasons or deliberately as we were not initially interested in blip data). The plot should not be considered a stability diagram, which would include a careful observation of the instability growth rates.

\section{Drift data and full model F predictions}\label{sec:appFdriftdata}

\begin{figure}
\centerline{\includegraphics[width=0.5\linewidth]{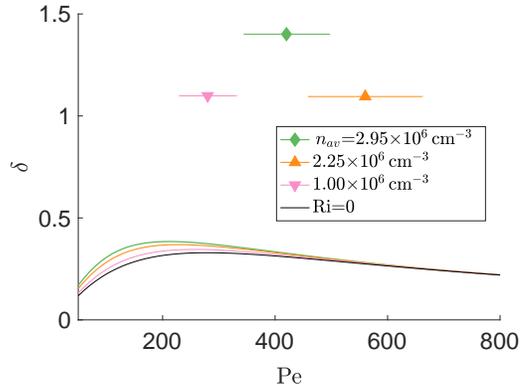}}
\caption{Predictions of model F for the fractional drift above the mean flow $\delta$ as a function of Pe. The theoretical curves are far from experimental drift values, as expected for a model that breaks down
at large Pe, see discussion in section \ref{sec:discussion}.}\label{fig:driftmodelF}
\end{figure}

The experimental values of the fractional drift above the mean flow $\delta$ (Figure \ref{fig:drift_thvsexp}) are provided in Table \ref{DriftValues} together with
theoretical values predicted using model G, with which they compare favourably (a similar quantitative comparison with model F is not worthwhile, as is clear from the inset of Figure \ref{fig:drift_thvsexp} and Figure \ref{fig:driftmodelF}). Bounds on the theoretical values were estimated by calculating the variations in the drift for upper and lower estimates of Pe and  $n_{av}$, i.e. $\Delta \delta_{\rm{Pe}} = \delta(\rm{Pe}+\Delta \rm{Pe}, n_{av})- \delta(\rm{Pe}, n_{av})$ and similarly for $\Delta \delta_{n_{av}}$. The resulting uncertainties were then combined by standard error propagation. The other experimental uncertainties have been previously discussed (those in the drift in section \ref{sec:methods}, while those in Pe in appendix \ref{sec:appCmicro}). 
The uncertainties in $n_{av}$ derive from the spectrophotometric calibration to concentration, discussed in section \ref{sec:methods}. 
\begin{table}
\begin{center}
\begin{tabular}{l l l l}
Pe & $\delta$ &   $\delta_{th}^G$ &  $n_{av} (10^6$ cells cm$^{-3})$\\
\hline
$280\pm50$ & $1.10\pm0.03$ &  $0.68^{+0.06}_{-0.08}$ &  $1.00\pm0.07$\\
$280\pm50$ & $2.02\pm0.03$ &  $1.93^{+2.51}_{-0.54}$ &  $3.65\pm0.27$\\
$420\pm76$ & $1.01\pm0.05$ &  $ 0.78\pm0.06$ &  $0.70\pm0.05$\\
$420\pm76$ & $1.40$ &  $1.34^{+0.45}_{-0.11}$ &  $2.95\pm0.22$\\
$560\pm101$ & $1.10\pm0.01$ &  $1.13\pm0.06$ &  $2.25\pm0.17$\\
\end{tabular} 
\caption{Experimental and theoretical values obtained using model G of the fractional drift compared, for the values of Pe and $n_{av}$ at which experiments were carried out. The evaluation of the uncertainties of the values shown is discussed in the text. No uncertainty is shown for experimental drift on the fourth line, as it was only possible to acquire one measurement of this drift (the single measurement error is negligible compared to the theoretical uncertainty estimates).} \label{DriftValues} 
\end{center}
\end{table}
As discussed in the text, the drift values predicted by model F are far from the experimentally observed values, whether reciprocal coupling due to buoyancy is included or not. This is clearly shown in Figure \ref{fig:driftmodelF}. 

\bibliographystyle{jfm}


\providecommand{\noopsort}[1]{}\providecommand{\singleletter}[1]{#1}%

\end{document}